\documentclass[12pt]{iopart}

\usepackage{iopams}  

\usepackage{graphicx}

\begin{document}

\title[Cloaking and anamorphism for light and mass diffusion]{Cloaking and anamorphism for light and mass diffusion}

\author{S\'ebastien Guenneau$^{1}$, Andr\'e Diatta$^{1}$, Tania M. Puvirajesinghe$^{2}$ and Mohamed Farhat$^{3}$}
\address{$^1$ Aix-Marseille Univ, CNRS, Centrale Marseille, Institut Fresnel, UMR CNRS 7249, Campus Universitaire de Saint-Jerome, 13013 Marseille, France
\\
Emails. sebastien.guenneau@fresnel.fr; ~andre.diatta@fresnel.fr.}
\address{$^2$ Aix-Marseille Univ, Institut Paoli-Calmettes, UMR INSERM 1068, UMR CNRS 7258, Marseille, France. Email. tania.guenneau-puvirajesinghe@inserm.fr
}
\address{$^3$ Qatar Environment and Energy Research Institute (QEERI), Hamad Bin Khalifa University, Qatar
Foundation, Doha, Qatar. Email. mfarhat@qf.org.qa}

\vspace{10pt}

\begin{abstract}
We first review classical results on cloaking and mirage effects for electromagnetic waves.
We then show that transformation optics allows the masking of objects or produces mirages in diffusive regimes.
In order to achieve this, we consider the equation for diffusive photon density in transformed coordinates, which is valid
for diffusive light in scattering media. More precisely, generalizing  transformations for star domains
introduced in [Diatta and Guenneau, J. Opt. 13, 024012, 2011] for matter waves, we numerically demonstrate that infinite conducting objects of
different shapes scatter diffusive light in exactly the same way.  We also propose a design of
external light-diffusion cloak with spatially varying sign-shifting parameters that hides a finite size scatterer outside the cloak.
We next analyse non-physical parameter in the transformed Fick's equation derived
in [Guenneau and Puvirajesinghe, R. Soc. Interface 10, 20130106, 2013], and propose to use a non-linear transform
that overcomes this problem. We finally investigate other form invariant transformed diffusion-like equations in the time domain,
and touch upon conformal mappings and non-Euclidean cloaking applied to diffusion processes.
\end{abstract}

\maketitle

\section{Introduction}
Following the 2006 proposal of an invisibility cloak via geometric transforms
in the covariant Maxwell's equations \cite{pendrycloak}, and their conformal counterpart in specific regimes \cite{ulf}, enhanced control of electromagnetic and plasmonic waves
has been experimentally validated in a number of fascinating studies \cite{schurig06,cai07,kante09,ulf08,guenneau10,spp1,spp2,spp3}, that also include acoustic \cite{cummer07,chen07,norris08}, hydrodynamic \cite{farhat08,alam12,maurel13,porter14,zareei15,dupont15,xu15,dupont16}, elastodynamic \cite{milton06,brun09,diatta14,farhat09,wegener12,colombi15,norris11,norris12,cummer16}
and even seismic \cite{brule14,colombi16} waves, see e.g. for a book devoted to these topics \cite{thebook}. Importantly, transformation optics \cite{pendrycloak} and conformal optics  \cite{ulf} have been
reconciled in the proposal of invisibility carpets \cite{li08}, that make use of quasi-conformal mappings that relax the constraint on anisotropy \cite{smith08,gabrielli09}.  

\begin{figure}[h]
\scalebox{1.0}{
\hspace{2cm}\mbox{}\includegraphics[width=12.5cm,angle=0]{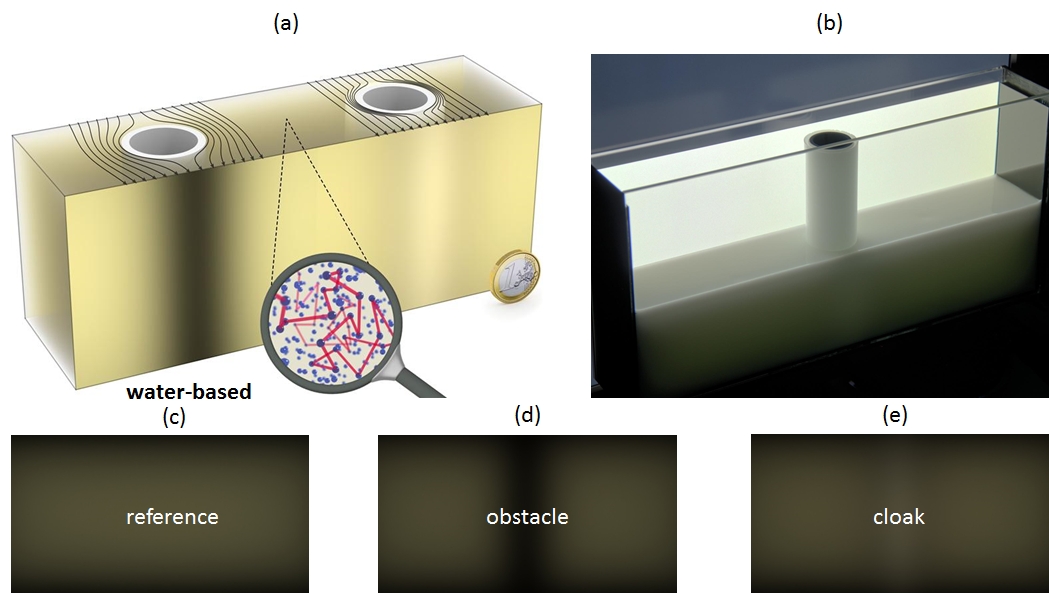}}
\mbox{}\vspace{-0.4cm}
\caption{Experiment on light diffusion cylindrical cloak by Wegener's group at KIT (figure courtesy of M. Kadic, see also \cite{wegener14}): (a) Illustration of the cloaking principle.
The streamlines visualize the photon-current-density vector field (the analog of the Poynting vector for ballistic light transport) as calculated in \cite{wegener14} from Fick's diffusion equation for homogeneous illumination for obstacle (left) and cloak (right);
The magnifying glass is an artistic microscopic view of light transport in diffusive media containing many randomly distributed scatterers.
(b) Photograph of the experimental setup with a cloaked obstacle; the diameters of the cylinders used are $2R_1 = 32.1$ mm and $2R_2 = 39.8$ mm and the tank thickness is $L = 60$ mm. (c) Experiment result (brightness) for light diffusion through water-based paint. (d) Same with an obstacle (note the darker region). (e) Same with a cloaked obstacle (brigthness is retrieved in comparison with (c) and (d)).
}\label{lightkit}
\end{figure}

Interestingly, control of wave trajectories can be extended to matter waves \cite{matter_prl2008}. Looking at the mathematical origin of cloaking, which
comes from the concept of Dirichlet-to-Neumann map in inverse problems with anisotropic media \cite{greenleaf03}, some of us proposed some mimetism
for matter waves \cite{diatta11}, which prompted further studies in control of diffusion phenomena that can be
achieved through coordinate transformations in the Fourier \cite{guenneau12} and Fick's \cite{guenneau13} equations for heat and mass respectively.
The latter work has served as an inspiration for the group of Martin Wegener, see Fig. \ref{lightkit}, who exploited the Fick's diffusion equation for multiple light scattering usages, whereby light is governed by ballistic laws \cite{wegener14}: cylindrical and spherical invisibility cloaks made of thin shells of polydimethylsiloxane doped with melamine-resin microparticles surrounding a diffusively reflecting hollow core were experimentally shown to achieve good cloaking performance in a water-based diffusive medium throughout the entire visible spectrum and for all illumination conditions and incident polarizations of light. However, these authors noted that the transformed Fick's equation as proposed in \cite{guenneau13} does not fully retain its form under Pendry's transform \cite{pendrycloak}, since a heterogeneous isotropic factor appears in front of the time derivative (a problem partially solved in \cite{guenneau13} using reduced parameters that preserve diffusion trajectories, but induce some impedance mismatch at the cloak's boundaries). This issue remained an Achilles heel in the control of mass diffusion through geometric transform until now. Interestingly, scattering cancellation techniques for mantle cloaking in diffusion processes do not suffer from this pitfall \cite{alu05,farhat15,farhat2016} as these are based on a totally different mechanism not underpinned by coordinate changes. This cloaking route reminiscent of neutral inclusions, a concept introduced by Kerner 60 years ago \cite{kerner56}, has been followed in a number of works in the theory of composites in the quasi-static limit \cite{bigoni98,milton02,farhat15a}. It can also explain the success of the aforementioned cloaking experiment in a water-paint based environment in \cite{wegener14}, see Fig. \ref{lightkit}, which has been extended to solid media invoking the concept of neutral inclusion, and no longer transformed Fick's equation, in \cite{schittny15}.

In this topical review paper, we focus our analysis on cloaking of diffusive light in scattering media via geometric transforms, which is governed by the diffusive photon density wave (DPDW) equation.
The transformed DPDW equation involves an anisotropic spatially varying diffusivity, as well as other heterogeneous but isotropic parameters. We shall first give a physical interpretation to all the terms in transformed DPDW. We will then consider numerical illustrations of anamorphism and external cloaking in DPDW equation. This part is actually an extension of \cite{farhat2016} where DPDW equation was first used to reduce scattering of diffuse light, however without resorting to coordinate changes. We will then consider a non-linear transform that makes Fick's equation totally form invariant in certain cylindrical geometries. We will finally come back to the interesting question of variation of the field inside the invisibility region in the time domain, since the linear transform which we use for Fick's equation avoids pitfall of a vanishing factor in front of the time derivative in transformed equations for the diffusion of mass, heat, matter and light. The concluding remarks are followed by perspectives on conformal mappings for diffusion processes and other parabolic equations of interest for geometric transforms.

\section{Transformed governing equations for diffusive light waves}
Let us set the scene of diffusion of light in scattering media.

\subsection{Introduction to the equation for propagation of diffuse photon density waves DPDW}

The transport equation for photons in scattering and absorbing media is given by \cite{case13,boas,oleary95}
\begin{equation}
\begin{array}{ll}
&\displaystyle{\frac{1}{\nu({\bf x})}\frac{\partial}{\partial t}L({\bf x},{\bf n},t) + {\bf n}\cdot\nabla\left(L({\bf x},{\bf n},t)\right) + (\mu_a+\mu_s^\prime)L({\bf x},{\bf n},t)} \\
&= \displaystyle{S({\bf x},{\bf n},t) + \mu_s^\prime \int dn'L({\bf x},{\bf n}',t) f({\bf n},{\bf n}')} \; ,
\end{array}
\label{diff1}
\end{equation}
with $L({\bf x},{\bf n},t)$ denoting the radiance at position ${\bf x}$, in direction ${\bf n}$, at time $t$ (with units of $Wm^{-2}sr^{-1}$). $f({\bf n},{\bf n}')$ denotes the probability of scattering to ${\bf n}'$ from direction ${\bf n}$. $\nu$ is the speed of light in the medium and $\mu_s^\prime$, $\mu_a$ are the scattering and absorption coefficients, respectively. $S$ is the source term. The photon fluence is given by:
\begin{equation}
\Phi({\bf x},t) = \int dn L({\bf x},{\bf n},t)  \; . \label{diff2}
\end{equation}
In order to get solutions of this equation for diverse geometries, one needs to make the so-called $P_N$ approximation \cite{furutsu94,frank07}. It consists in expanding the radiance $L$ and source $S$ in terms of spherical harmonics $Y_{l,m}$, and truncating at order $l=N$, i.e.
\begin{equation}
L({\bf x},{\bf n},t) = \sum_{l=0}^N\sum_{m=-l}^l\phi_{l,m}({\bf x},t)Y_{l,m}({\bf n})  \; , \label{diff3}
\end{equation}
and
\begin{equation}
S({\bf x},{\bf n},t) = \sum_{l=0}^N\sum_{m=-l}^ls_{l,m}({\bf x},t)Y_{l,m}({\bf n})  \; . \label{diff4}
\end{equation}
The phase function can also be expanded as (assuming that the scattering amplitude is only dependent on the change in direction of the photon \cite{furutsu94}):
\begin{equation}
f({\bf n},{\bf n}')=\sum_{l=0}^N\sum_{m=-l}^l g_lY_{l,m}^*({\bf n}')Y_{l,m}({\bf n}) \; .
\label{phase}
\end{equation}
The approximation $P_1$ (i.e. setting $N=1$) is actually, quite appropriate when $\mu_a\ll\mu_s^\prime$ (absorption much smaller than scattering) and $f$ is not too anisotropic  \cite{oleary95,furutsu94,frank07}. Inserting Eqs.~(\ref{diff3})-(\ref{diff4}) and the development of $f$ in Eq.~(\ref{diff1}) yields, under the $P_1$ approximation
\begin{eqnarray}\label{diff5}
-\nabla\cdot \left(D({\bf x}) \nabla \Phi({\bf x},t)\right) + \mu_a({\bf x})\Phi({\bf x},t) + \frac{1}{\nu({\bf x})} \frac{\partial \Phi({\bf x},t)}{\partial t} - S_0 = \\ 
- \frac{3D({\bf x})}{\nu({\bf x})}\left[\mu_a({\bf x})\frac{\partial\Phi({\bf x},t)}{\partial t}+\frac{1}{\nu({\bf x})}\frac{\partial^2\Phi({\bf x},t)}{\partial t^2}\right] + \frac{3D({\bf x})}{\nu({\bf x})}\frac{\partial S_0}{\partial t} - 3D({\bf x}) \nabla\cdot S_1 \; \nonumber , 
\end{eqnarray}
with $D$ the modified diffusivity, i.e. $D=1/(3\mu_s)$ and $\mu_s$ the modified scattering coefficient $\mu_s=\mu_s^\prime(1-g_1)$, where $g_1$ depends on the anisotropy of $f$, see (\ref{phase}), and often has a value close to $0.99$. In the frequency domain, i.e. assuming a time dependence of $\Phi\propto e^{-i\omega t}$ and assuming further that the scattering frequency is much larger than $\omega$, i.e. $\nu\mu_s\gg\omega$, one can finally get (after ignoring the dipole term of the source) the photon diffusion equation by setting the right side of Eq.~(\ref{diff5}) equal to zero, so-called DPDW equation, that we will utilize throughout this paper, namely:
\begin{equation}
-\nabla\cdot \left(D({\bf x}) \nabla \Phi({\bf x},t)\right) + \frac{1}{\nu({\bf x})}\frac{\partial\Phi({\bf x},t)}{\partial t} + \mu_{a}({\bf x}) \Phi({\bf x},t) = S_0 \; . \label{govpressuret}
\end{equation}

This equation for propagation of diffuse photon density waves with heterogeneous isotropic diffusivity, and heterogeneous absorption coefficient, in the time-harmonic regime, takes thus the final form
\begin{equation}
-\nabla\cdot \left(D({\bf x}) \nabla \Phi({\bf x})\right) - \frac{i\omega}{\nu({\bf x})}\Phi({\bf x}) + \mu_{a}({\bf x}) \Phi({\bf x}) = S_0 \; . \label{govpressure}
\end{equation}
where $\Phi$ represents the distribution of fluence field at each point ${\bf x}=(x,y,z)$ in Cartesian
coordinates (for more details on cylindrical DPDW cloaks in the frequency domain, see \cite{farhat2016}).

Moreover, let us recall that $D$ is the modified diffusivity, i.e. $D=1/(3\mu_s)$, with $\mu_s$ the
modified scattering coefficient and $\nu$ is the speed of light as defined in (\ref{diff1}).

\noindent Let us now consider a map (a pull-back) from a co-ordinate system
$\{x',y',z'\}$ to the co-ordinate system $\{x,y,z\}$ given by the
transformation characterized by $x(x',y',z')$, $y(x',y',z')$ and
$z(x',y',z')$. This change of co-ordinates is characterized by the
transformation of the differentials through the Jacobian:

\begin{equation}
\left(%
\begin{array}{c}
  dx \\
  dy \\
  dz \\
\end{array}%
\right) = \mathbf{J}_{xx'}
\left(%
\begin{array}{c}
  dx' \\
  dy' \\
  dz' \\
\end{array}%
\right) \; , \hbox{with} \; \mathbf{J}_{xx'}=
\frac{\partial(x,y,z)}{\partial(x',y',z')} \;.
\end{equation}

\noindent From a geometric point of view, the matrix $\mathbf{T} \!=
\! \mathbf{J}_{xx'}^T \mathbf{J}_{xx'}/\det(\mathbf{J}_{xx'})$ allows for a description of the metric tensor inside the cloak. The only method of operation for the transformed
coordinates is to replace the diffusivity (homogeneous and
isotropic) and potential by equivalent ones. The effective diffusivity
becomes heterogeneous and anisotropic, while the absorption coefficient
and the speed of light deserve a new expression. Their properties are given by
\begin{equation}
\underline{\underline{D'}} =D \mathbf{T}_T^{-1} \; ,
\; \nu'={T}_{zz} \nu \; ,
\; \mu_{a}'={T}_{zz}^{-1} \mu_{a} \; ,
\; S'={T}_{zz}^{-1} S \; ,
\label{epsmuT}
\end{equation}
where $\mathbf{T}_T^{-1}$ stands for the upper diagonal part of the
inverse of $\mathbf{T}$ and $T_{zz}$ is the third diagonal entry of
${\bf T}$.

\noindent The transformed equation associated with
(\ref{govpressure}) reads
\begin{equation}
-\nabla\cdot \left(\underline{\underline{D'}}({\bf x'}) \nabla \Phi({\bf x'})\right) - \frac{i\omega}{\nu'({\bf x'})}\Phi({\bf x'}) + \mu'_{a}({\bf x'}) \Phi({\bf x'}) = S'({\bf x'}) \Phi({\bf x'})
\; , \label{transfpressure}
\end{equation}
where importantly the source term $S'$ is a hallmark of the geometric transform. Indeed, if one places a source in the transformed medium, it seems to radiate from a shifted
location, what can be considered as a mirage effect \cite{zolla07}.

\noindent An elegant way to derive (\ref{transfpressure}) is to multiply (\ref{govpressure}) by a test function that is an infinitely differentiable function which vanishes on the boundary $\partial\Omega$ of a domain $\Omega$, and to further integrate by parts over $\Omega$, which leads to \footnote{We have used the fact that $\int_{\Omega} d{\bf x}  \nabla \cdot\left( (D \nabla \Phi) {\phi} \right)=\int_{\Omega} d{\bf x} (D \nabla \Phi) \cdot  \left( \nabla {\phi} \right)
+\int_{\Omega} d{\bf x} {\phi} \nabla\cdot\left( D \nabla \Phi \right)$ and we invoked the divergence theorem in the left hand side of the equation, which esnures that this term is equal to
$\int_{\partial\Omega} ds\left(D \nabla \Phi \cdot {\bf n}{\phi}\right)$, with ${\bf n}$ the unit outward normal to the boundary $\partial\Omega$.}:
\begin{equation}
\begin{array}{lll}
&\displaystyle{\int_{\Omega} d{\bf x}\left( \frac{-i\omega+\mu_a\nu}{\nu} \Phi\phi \right)
+\int_{\Omega} d{\bf x} \left( D \nabla \Phi \cdot \nabla {\phi} \right)} \nonumber \\
&\displaystyle{-\int_{\partial\Omega} ds\left(D \nabla \Phi \cdot {\bf n}{\phi}\right)  + <S,\phi>}
 = 0 \; ,
\end{array}
\label{equ_ac_weak}
\end{equation}
where ${\bf n}$ is the unit outward normal to the boundary $\partial\Omega$ of the integration domain
$\Omega$. Moreover, $<,>$ denotes the duality product between the space of Distributions
(these are generalized functions such as Dirac, Heaviside functions) and the space of test functions
(i.e. infinitely differentiable functions with compact support on $\Omega$)
i.e. a pairing in which one integrates a distribution against a test function.

\noindent We now apply to (\ref{equ_ac_weak}) the coordinate change ${\bf x}=(x,y,z) \rightarrow {\bf x'}=(x',y',z')$
(a push-forward)
and noting that
\begin{equation}
\begin{array}{ll}
&\nabla= (\partial/\partial x,\partial/\partial y,\partial/\partial z)^T= \displaystyle{\frac{\partial(x',y',z')}{\partial(x,y,z)} (\partial/\partial x',\partial/\partial y',\partial/\partial z')^T} \nonumber \\
&={\bf J}_{x'x} \nabla'= {\bf J}_{xx'}^{-1} \nabla'
\end{array}
\end{equation}
where $\nabla'$ is the gradient in the new coordinates, we end up with
\begin{equation}
\begin{array}{ll}
&\displaystyle{\int_{\Omega}d{\bf x'}\left( \frac{-i\omega+\mu_a\nu}{\nu} \Phi\phi  \hbox{det}({\bf J}_{xx'}) \right)} \nonumber \\
&+\displaystyle{\int_{\Omega}d{\bf x'}\left\{  \left( {\bf J}_{xx'}^{-1} \nabla' {\phi} \cdot D
{\bf J}_{xx'}^{-1} \nabla' \Phi \right) \hbox{det}({\bf J}_{xx'})\right\}}\\
&-\displaystyle{\int_{\partial\Omega} ds' \left(D {\bf J}_{xx'}^{-1}\nabla' \Phi \cdot {\bf n}{\phi}\right)\hbox{det}({\bf J}_{xx'})} \nonumber \\
&+< \hbox{det}({\bf J}_{xx'})S,\phi>
= 0 \; .
\end{array}
\label{disappear1}
\end{equation}
Upon integration by parts,
and noting that $ {\bf J}_{xx'}^{-1} \nabla' {\phi} \cdot D
{\bf J}_{xx'}^{-1} \nabla' \Phi={(\nabla' {\phi})}^T {\bf J}_{xx'}^{-T}
D{\bf J}_{xx'}^{-1} \nabla' \Phi$
we obtain:
\begin{equation}
\begin{array}{ll}
&\displaystyle{\int_{\Omega}d{\bf x'}\left( \frac{-i\omega+\mu_a\nu}{\nu} \Phi\phi  \hbox{det}({\bf J}_{xx'}) \right)} \nonumber \\
&-\displaystyle{\int_{\Omega}d{\bf x'}\left\{  \nabla' \cdot \left( {\bf J}_{xx'}^{-1}D{\bf J}_{xx'}^{-T} \nabla' {\Phi} \right) \phi \hbox{det}({\bf J}_{xx'})\right\}}\\
& +< \hbox{det}({\bf J}_{xx'})S,\phi>
= 0 \; ,
\end{array}
\label{disappear2}
\end{equation}
which is the integral form of (\ref{transfpressure}). This lays the foundation of transformation light diffusion \footnote{The boundary integral disappears between (\ref{disappear1}) and (\ref{disappear2}), which should not be a surprise since this second integration by parts is just the opposite of the one performed before coordinate change, see (\ref{equ_ac_weak}), and in any case one could simply consider test functions $\phi$ that vanish on the boundary of $\Omega$.}.

Note that in the case of cylindrical domains, when the geometric transform only affects the transverse coordinates
i.e. the transformation is characterized by $x(x',y')$, $y(x',y')$, $z(z')$ one has:
\begin{equation}
\begin{array}{ll}
&\underline{\underline{D'}} =D \mathbf{T}_T^{-1} \; ,
\; \nu'=\hbox{det}({\bf J}_{xx'})^{-1}\nu={T}_{zz} \nu \; , \nonumber \\
&\; \mu_{a}'=\hbox{det}({\bf J}_{xx'})\mu_{a}={T}_{zz}^{-1} \mu_{a} \; ,
\; S'={T}_{zz}^{-1} S \; ,
\end{array}
\label{epsmuT}
\end{equation}
where $\mathbf{T}_T^{-1}$ stands for the upper diagonal part of the
inverse of $\mathbf{T}$ and $T_{zz}$ is the third diagonal entry of
${\bf T}$. This is left as an exercise for the zealous reader, who can otherwise find the derivation
in section 4.5. on external cloaking.

\begin{figure}[h]
\scalebox{1.0}{
\hspace{2cm}\mbox{}\includegraphics[width=12.5cm,angle=0]{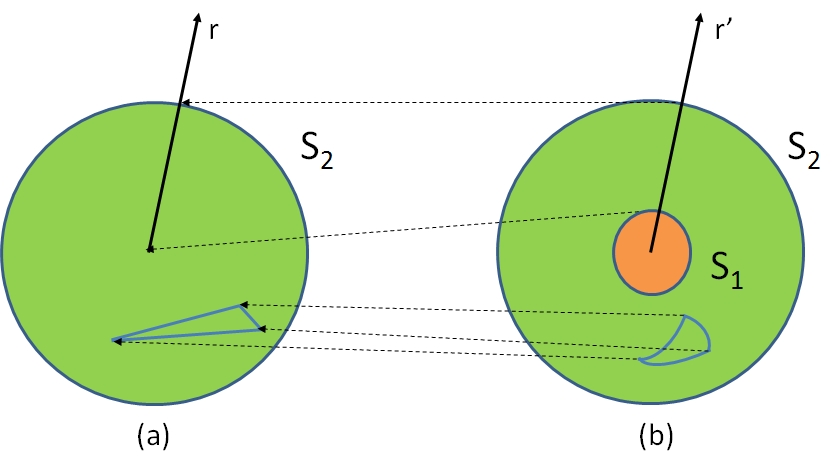}}
\mbox{}\vspace{-0.4cm}
\caption{Construction of singular cloak for anamorphism. Pendry's transformation
$r'=f(r)=r_1+r(r_2-r_1)/r_2,$~~$\theta'=\theta,$~~$z'=z,$ with inverse $r=f^{-1}(r')$,
~~$\theta=\theta',$~~$z=z',$ shrinks a cylindrical region of radius $r_2$
bounded by the surface $S_2$ as in (a),
into a hollow cylindrical region bounded by $S_1$ and $S_2$ with inner  and outer $r_1$ and $r_2,$
respectively  as in (b).
The curvilinear metric inside the cloak is
described by the transformation matrix ${\bf T}^{-1}={\bf J}_{xx'}^{-1}{\bf J}_{xx'}^{-T}\hbox{det}({\bf J}_{xx'})$
where ${\bf J}_{xx'}^{-T}$ is the transpose of the inverse ${\bf J}_{xx'}^{-1}$ of
${\bf J}_{xx'}={\bf J}_{xr}{\bf J}_{rr'}{\bf J}_{r'x'},$ $\;\; {\bf J}_{xr}=\frac{\partial(x,y,z)}{\partial(r,\theta,z)}$
 and $\mathbf{J}_{rr'}=\frac{\partial(r,\theta,z)}{\partial(r',\theta',z')}$.
Physically, an object with diffusivity $D{\bf T}^{-1}$ and a boundary ${\bf x}'(t)$ placed
in the diffusion cloak (b) scatters like an object of diffusivity $D$ placed
in a homogenous medium of diffusivity $1$ and a boundary ${\bf x}(t)$ (Cartesian coordinates)
with the same variation of the parameter $t>0$. If one would like to design an object with
diffusivity $D$ and boundary ${\bf x}(t)$, it is best to consider the push-forward
${\bf x}'(t)=\left((r_2-r_1)/r_2+r_1/\Vert {\bf x}(t) \Vert\right){\bf x}(t)$, with $\Vert
{\bf x}(t) \Vert=r$. This route  \cite{nicolet10} that generalizes cloaking and
masking \cite{teixeira07} via differential geometry was coined anamorphism by
Nicolet et al. \cite{nicolet10}, and this served as an inspiration for other research groups \cite{cui2,cui3,amra}. Note that an inverse mimicking algorithm has been proposed in \cite{amra}.}\label{geojohn}
\end{figure}

We would like to also stress that we shall use pull-back transformations in the sequel for historical reasons:
cloaking was first proposed for Maxwell's equations, which behave appropriately under pull-back geometric transforms,
unlike push-forward \cite{zolla07}, and this makes possible designs of arbitrarily shaped cloaks \cite{nicolet08a,cui1,dupont09}. Let us also take this opportunity to give proper credit to the seminal
work on masking by Teixeira \cite{teixeira07} and on anamorphism by Nicolet, Zolla and Geuzaine \cite{nicolet10},
that served as an inspiration for other research groups \cite{cui2,cui3,cui4,amra}.
In the former, the formalism of differential geometry is used to demystify cloaking and masking, while the latter shows
that if an object is placed in a transformed medium, it will scatter like another object with deformed
boundary, and different electromagnetic parameters, see Fig. \ref{geojohn}, which is characterized by the pull-back
transform. Nicolet et al. illustrated their theory with an object placed inside a
heterogeneous anisotropic shell, which is a singular cloak. In what follows, we
shall pursue an alternative route to anamorphism, which amounts to placing
an object in the invisibility region of a non-singular cloak that scatters like a different object,
see \cite{diatta11} for the case of matter-waves.

\section{Transformation Optics for star domains}\label{chap:transformation}
In this section we discuss a generalization of Pendry's linear transformation \cite{pendrycloak}. It is applied to the general framework  where it sends a domain to another domain, and is nowhere singular. The corresponding derived electromagnetic material is hence nowhere singular. Moreover, this allows the creation of mimesis whereby an object dressed with a device acquires the same electromagnetic behavior as another one of our choice. The case where the polynomial is of degree $1$ is discussed in \cite{diatta11}.

\subsection{Governing equation for heat}
We consider the anisotropic diffusion equation  in a two-dimensional domain without heat source
\begin{equation}
\rho({\bf x}) c({\bf x})\frac{\partial u}{\partial t}=\nabla\cdot (\underline{\underline{\kappa}}({\bf x})\nabla u)
\; ,
\label{heat1}
\end{equation}
where $u$ represents the distribution of temperature evolving with time $t>0$,  at each point ${\bf x}=(x,y)$ in the domain.
Moreover, the thermal conductivity $\underline{\underline{\kappa}}$ ($W.m^{-1}.K^{-1}$ i.e. watt per meter kelvin in SI units) has the form
\begin{eqnarray}
\underline{\underline{\kappa}} = \left(
\begin{array}{ccc}
\kappa_{11} &\kappa_{12} \cr
\kappa_{21} &\kappa_{22}
\end{array}
\right)
\end{eqnarray}
with $\kappa_{12}=\kappa_{21}$.
Also, $\rho$ is the density ($kg.m^{-3}$ i.e. kilogram per cubic meter in SI units)
and $c$ the specific heat (or thermal) capacity ($J.K^{-1}.kg^{-1}$ i.e. joule per kilogram kelvin in SI units).

\subsection{Governing equations for transverse electromagnetic waves}
Let us now consider the Maxwell's equations in a cylindrical domain without source.
Let us also assume that the medium is described by anisotropic
permittivity $
\underline{\underline{\varepsilon}}$ and permeability $\underline{\underline{\mu}}$ that have the form
\begin{eqnarray}
\underline{\underline{\varepsilon}}({x,y}) = \left(
\begin{array}{ccc}
\varepsilon_{11} &\varepsilon_{12}        & 0 \cr
\varepsilon_{21} &\varepsilon_{22}   & 0 \cr
 0      & 0        & \varepsilon_{33}
\end{array}
\right)
\; , \;
\underline{\underline{\mu}}({x,y}) = \left(
\begin{array}{ccc}
\mu_{11} &\mu_{12}        & 0 \cr
\mu_{21} &\mu_{22}   & 0 \cr
 0      & 0        & \mu_{33}
\end{array}
\right)
\; ,
\end{eqnarray}
with $\varepsilon_{12}=\varepsilon_{21}$ and $\mu_{12}=\mu_{21}$.

In transverse electric polarisation, the Maxwell operator writes as
\begin{eqnarray}\label{eq:Hl}
\nabla\times\left(
\underline{\underline{\varepsilon}}^{-1}({x,y})\nabla\times {\bf H}_l
\right) = -\frac{\partial^2}{\partial t^2}
\left(\underline{\underline{\mu}}({x,y}){\bf H}_l \right)
\end{eqnarray}
where ${\bf H}_l=H_z(x,y,t){\bf e}_z$
and in transverse magnetic polarisation it writes as
\begin{eqnarray}\label{eq:El}
\nabla\times\left(
\underline{\underline{\mu}}^{-1}(x,y)\nabla\times {\bf E}_l
\right) = -\frac{\partial^2}{\partial t^2}
\left(\underline{\underline{\varepsilon}}(x,y){\bf E}_l\right)
\end{eqnarray}
where ${\bf E}_l=E_z(x,y,t){\bf e}_z$.

We would like to show that one can recast (\ref{eq:Hl}) and
(\ref{eq:El}) in the form of scalar wave equations. For this,
we need the following result:

{\bf Property}: Let ${\bf M}$ be a real symmetric matrix defined as
follows
\begin{eqnarray}
{\bf M} = \left(
\begin{array}{ccc}
 m_{11} & m        & 0 \cr
 m      & m_{22}   & 0 \cr
 0      & 0        & m_{33}
\end{array}
\right) = \left(
\begin{array}{ccc}
{\bf M}_T & 0 \cr
 0         & m_{33}
\end{array}
\right) \; .
\end{eqnarray}

Then we have
\begin{equation}
\nabla\times \Biggl ( {\bf M}\nabla\times \Bigl ( u(x,y)
\mathbf{e}_z \Bigr ) \Biggr )= - \nabla\cdot \Biggl ( {\bf M}_T^{-1} \hbox{det}({\bf M}_T)\nabla u(x,y) \Biggr
)\mathbf{e}_z \; .
\end{equation}

Indeed, we note that
\begin{equation}
\begin{array}{ll}
\nabla\times \Biggl ( {\bf M} \nabla\times \Bigl ( u(x,y)
\mathbf{e}_z \Bigr ) \Biggr )
&= \displaystyle{- \Biggl ( \frac{\partial}{\partial x} \Bigl (
m_{22} \frac{\partial u}{
\partial x} - m \frac{\partial u}{\partial y} \Bigr )}  \nonumber\\
&\displaystyle{+ \frac{\partial}{\partial y} \Bigl ( m_{11}
\frac{\partial u}{\partial y} - m \frac{\partial u}{\partial x}
\Bigr ) \Biggr ) \mathbf{e}_z} \; .
\end{array}
\label{trick}
\end{equation}

Furthermore, let ${\bf M}'$ be defined as
\begin{eqnarray}
{\bf M}' = \left(
\begin{array}{ccc}
 m'_{11}      & m'_{12} \cr
 m'_{21}      & m'_{22}
\end{array}
\right) \; .
\end{eqnarray}
From (\ref{trick}), we have $$\nabla\times \Bigl ( {\bf M} \nabla\times \Bigl ( u(x,y)
\mathbf{e}_z \Bigr ) \Biggr ) = - \nabla\cdot \Biggl ( {\bf M}'
\nabla u \Biggr )\mathbf{e}_z \; ,$$ if and only if
$$m'_{11} \frac{\partial u}{\partial x} + m'_{12} \frac{\partial u}{\partial y}
=m_{22} \frac{\partial u}{\partial x} - m \frac{\partial u}{\partial y}$$
$$m'_{21} \frac{\partial u}{\partial x} + m'_{22} \frac{\partial u}{\partial y}
=- m \frac{\partial u}{\partial x} + m_{11} \frac{\partial u}{\partial y}
\; ,$$
which is true if ${\bf M}' = {\bf M}_T^{-1}\hbox{det}({\bf M}_T)$.

\noindent Using the above property, from (\ref{eq:Hl}), we derive
that
\begin{eqnarray} \nabla\cdot
\left[\frac{1}{\varepsilon_{11}\varepsilon_{22}-\varepsilon_{12}\varepsilon_{21}}
\left(
\begin{array}{ccc}
\varepsilon_{11}      & \varepsilon_{12} \cr
\varepsilon_{21}      & \varepsilon_{22}
\end{array}
\right)  \nabla H_z \right]
= \frac{\partial^2}{\partial t^2} \left(\mu_{33} H_z\right)
\; ,
\label{eq:Hl1}
\end{eqnarray}
and from (\ref{eq:El}), we derive
that
\begin{eqnarray} \nabla\cdot
\left[\frac{1}{\mu_{11}\mu_{22}-\mu_{12}\mu_{21}}
\left(
\begin{array}{ccc}
\mu_{11}      & \mu_{12} \cr
\mu_{21}      & \mu_{22}
\end{array}
\right)  \nabla E_z \right]
= \frac{\partial^2}{\partial t^2} \left(\varepsilon_{33} E_z\right)
\; .
\label{eq:El1}
\end{eqnarray}
We note that for isotropic permittivity and permeability, (\ref{eq:Hl1}) and (\ref{eq:El1})
reduce to the usual expressions.

If we were to solve the heat equation (\ref{heat1}) in anisotropic media
using some electromagnetic code, (\ref{eq:Hl1}) and (\ref{eq:El1})
might be useful  for instance in a commercial finite element package like COMSOL.

\subsection{Construction of a star cloak via non-linear transform} 
The non-linear transformation  (\ref{eqnonlinear}) we propose here readily extends to
 any star domain in $\mathbb R^p,$ that is,  any domain  with a vantage point from which all points, including the boundary's,  are
 within line-of-sight.  The transformation preserves every line passing through
 that fixed vantage point. In more details, consider three bounded star domains
   $D_j$ in $\mathbb R^p, ~ p=2,3$ with piecewise smooth boundaries  ${\mathbf S}_j:=\partial D_j,$ all sharing the
    same chosen vantage point, $j=0,1,2,$  and further suppose they have the following configuration: $D_2$ contains $D_1$ which in turn, contains $D_0.$
    We would like to make  the domain $D_2$  acquire the same response to any electromagnetic waves  as  the different domain $D_0.$  More precisely, we seek to fill the region  $D_2\setminus D_1$ with an electromagnetic medium whose properties (permittivity and permeability) 
are given by the tensor ${\mathbf T}^{-1}$  in Equation (\ref{eq:tensorgeneral}) below, so that  $D_2$ appears to electromagnetic waves as if  it were $D_0$ with an infinitely conducting boundary ${\mathbf S}_0:=\partial D_0.$

The transformation (\ref{eqnonlinear}) is  the identity
      outside $D_2$, that is, in $\mathbb R^p\setminus D_2$. It sends the region
      $D_2\setminus D_0$  to the region $D_2\setminus D_1$, in such a way that the boundary
       ${\mathbf S}_2:=\partial D_2$ of $ D_2$ stays point-wise fixed, while ${\mathbf S}_0$ is mapped to
       ${\mathbf S}_1:=\partial D_1.$
  We also set  an infinitely conducting boundary condition on the inner boundary ${\mathbf S}_0$  of the region $D_2\setminus D_1.$ Any type of defect could be concealed inside $D_1$ and  $D_2$ will yet still have
the same electromagnetic response as $D_0$ with an infinitely conducting boundary ${\mathbf S}_0.$
In practice, we may divide the domains $D_j$ into subdomains, the part of whose boundaries lying inside ${\mathbf S}_j$ is a smooth hypersurface.

\begin{figure}[h]
\scalebox{1.0}{
\hspace{2cm}\mbox{}\includegraphics[width=12.5cm,angle=0]{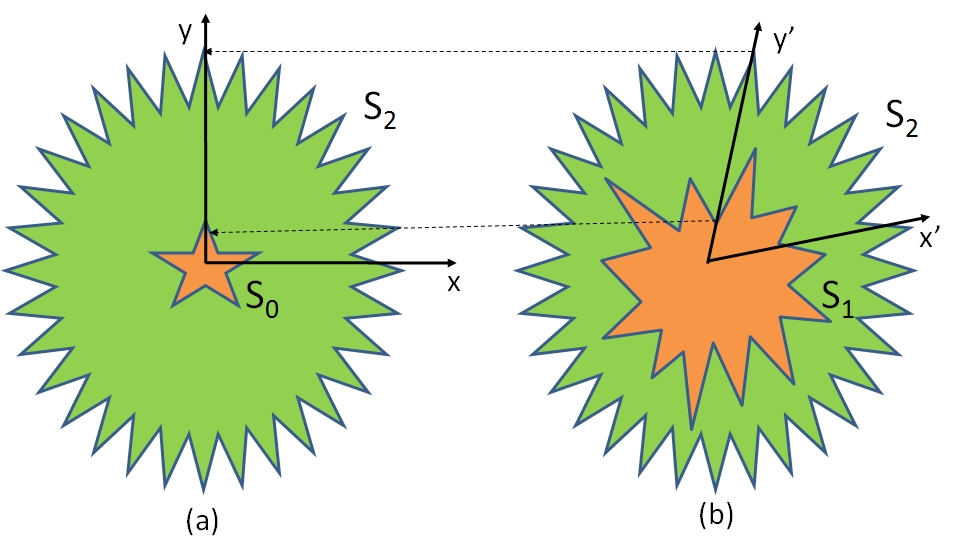}}
\mbox{}\vspace{-0.4cm}
\caption{Construction of non-singular cloaks allowing mimetism. The transformation  with inverse (\ref{eq:cylinders}) shrinks the
region bounded by the two surfaces $S_0$ and
${\mathbf S}_2$ (a) into the region bounded by ${\mathbf S}_1$ and ${\mathbf S}_2$ (b). The curvilinear metric inside the carpet is
described by the transformation matrix ${\bf T},$ see (\ref{eq:tensorgeneral})-(\ref{eq:partialderivyy}).
Physically, an object with Dirichlet or Neumann data on boundary $S_1$ scatters like another object with Dirichlet or Neumann data on boundary $S_0$, which was coined mimetism in \cite{diatta11}. Importantly, if one considers a source $S'$ with support on $S_1$ (e.g. generated by a current), then it will behave like a source $S$ with support on $S_0$. Combining the concept of anamorphism in Fig. \ref{geojohn} with mimetism, can be achieved by considering an object of anisotropic heterogeneous diffusivity in $S_1$ that would behave like another object with different anisotropic heterogeneous diffusivity, which is similar to anamorphism and mimetism.}\label{geocloak}
\end{figure}

Here is a detailed description of the  non-linear transformation and the subsequent construction of the tensor ${\mathbf T^{-1}}$. Consider a point ${\bf \underline x}$ of $D_1\setminus D_0$ with ${\bf \underline x}=(x^1, x^2,...)$ relative to a system of coordinates centered at the chosen  vantage point {\bf \underline 0}. The line passing through {\bf \underline x} and  {\bf \underline 0} meets the boundaries  ${\mathbf S}_0,$ ${\mathbf S}_1,$ ${\mathbf S}_2$ at the unique points ${\bf \underline x_0}=(x_0, y_0,...),$  ${\bf \underline x_1}=(x_1, y_1,...)$,  ${\bf \underline x_2}=(x_2, y_2,...)$ respectively. Inside 
$D_2\setminus D_0$, the transformation reads:

\begin{eqnarray}
\left\{
\begin{array}{lr}
x'= {(\alpha^{-1} {(x-x_0)}^n+x_1^n)}^{1/n}, ~ \rm{ with } ~\alpha=\frac{{(x_2-x_0)}^n}{x_2^n-x_1^n}
\\
y'= {(\beta^{-1} {(y-y_0)}^n+y_1^n)}^{1/n} ~ \rm{ with }~ \beta=\frac{{(y_2-y_0)}^n}{y_2^n-y_1^n}
\\
z'= {(\gamma^{-1} {(z-z_0)}^n+z_1^n)}^{1/n}, ~ \rm{ with } ~\gamma=\frac{{(z_2-z_0)}^n}{z_2^n-z_1^n}
\end{array}
\right.
\label{eqnonlinear}
\end{eqnarray}
where $n$ is a strictly positive integer (note that the linear case whereby $n=1$ was studied in \cite{diatta11}).

For the practical  implementation, we actually need the inverse ${\bf \underline x'} \mapsto {\bf \underline x}$  of (\ref{eqnonlinear}) which is the general setting that reads $x^i=x_0^i+\alpha_i~ ({(x'^i)}^n-{(x^i_1)}^n), ~ \rm{ where } ~\alpha_i=\frac{{(x_2^i-x^i_0)}^n}{{(x_2^i)}^n-{(x^i_1)}^n}.$ In  the 3-space with coordinates $(x^1, x^2,x^3)=(x,y,z)$ we can write it as
\begin{eqnarray}
\left\{
\begin{array}{lr}
x=x_0+{[\alpha~ (x'^n-x_1^n)]}^{1/n}, ~ \rm{ with } ~\alpha=\frac{{(x_2-x_0)}^n}{x_2^n-x_1^n}
\\
y=y_0+{[\beta~(y'^n-y_1^n)]}^{1/n} ~ \rm{ with }~ \beta=\frac{{(y_2-y_0)}^n}{y_2^n-y_1^n}
\\
z=z_0+{[\gamma~ (z'^n-z_1^n)]}^{1/n}, ~ \rm{ with } ~\gamma=\frac{{(z_2-z_0)}^n}{z_2^n-z_1^n}.
\\
\end{array}
\right.
\end{eqnarray}

The boundary points  ${\mathbf x_0}=(x_0, y_0,...),$  ${\mathbf  x_1}=(x_1, y_1,...)$,  ${\mathbf  x_2}=(x_2, y_2,...)$  are functions of $x,y,\cdots$ and hence of $x',y',\cdots$ We will discribe them explicitly in Sections \ref{chap:linearboundaries}, \ref{chap:squarecircle} , for some particular cases.


The tensor ${\bf T}^{-1}$ below gives the expression of the electromagnetic properties of the transformed material inside $D_2\setminus D_1.$ 
Indeed, denote the permittivity and the permeability of the vacum by $\varepsilon_0$ and $\mu_0$ respectively. Then one derives the permittivity and the permeability of the new electromagnetic transformed material obtained using transformation  (\ref{eq:cylinders}), via the formula
\footnote{If we replace vaccuum by an anisotropic media, one has $\underline{\underline{\varepsilon}}'(\mathbf x')
= {\bf J}_{xx'}^{-1} {\underline{\underline{\varepsilon}}(\mathbf x)}
{\bf J}_{xx'}^{-T}\hbox{det}({\bf J}_{xx'})$
and
$\underline{\underline{\mu}}'(\mathbf x')= {\bf J}_{xx'}^{-1} {\underline{\underline{\mu}}(\mathbf x)} {\bf J}_{xx'}^{-T}\hbox{det}({\bf J}_{xx'})$ instead of (\ref{nicolet}).}
\begin{equation}
{\bf \varepsilon }:=\varepsilon_0 {\bf T}^{-1} \; , \; {\bf \mu }:=\mu_0 {\bf T}^{-1} \; ,
\label{nicolet}
\end{equation}
where $\epsilon_0\mu_0=c^{-2}$, with $c$ the speed of light in vacuum.

In this article, we focus on light diffusion phenomena, but the above permittivity and permeability tensors (the square root of their product can be encapsulated within an anisotropic spatially varying refractive index) simply translate into diffusivity, scattering and absorption coefficients. Therefore and we consider the cases where  the regions $D_j$ are cylinders over some plane curves (triangular, square, elliptic, sun flower-like cylinders, etc.) Thus, we consider the transformation mapping the region enclosed between the cylinders ${\bf S}_0$ and  ${\bf S}_2$ into the space between  ${\bf S}_1$ and  ${\bf S}_2$ as in Fig. \ref{geocloak}, whose inverse is
 \begin{eqnarray}
\left\{
\begin{array}{lr}
x=x_0+{[\alpha~ (x'^n-x_1^n)]}^{1/n}, ~ \rm{ with } ~\alpha=\frac{{(x_2-x_0)}^n}{x_2^n-x_1^n}
\\
y=y_0+{[\beta~(y'^n-y_1^n)]}^{1/n} ~ \rm{ with }~ \beta=\frac{{(y_2-y_0)}^n}{y_2^n-y_1^n}
\\
z=z'
\end{array}
\right.\label{eq:cylinders}
\end{eqnarray}


The tensor ${\bf T}^{-1}$ is thus given by
\begin{equation}
{\bf T}^{-1}=
\left(
\begin{array}{lll}
T^{-1}_{11}&T^{-1}_{12}&0\\
T^{-1}_{12}&T^{-1}_{22} &0\\
0&0&T^{-1}_{33}
\end{array}
\right)
\label{eq:tensorgeneral}
\end{equation}
with
\begin{eqnarray} T^{-1}_{11}=\frac{a_{11}}{a_{33}}, ~~T^{-1}_{12}=\frac{a_{12}}{a_{33}}, ~ ~ T^{-1}_{22}=\frac{a_{22}}{a_{33}},~~
T^{-1}_{33}=a_{33},\label{eq:coefgeneral1}\end{eqnarray} where the $a_{ij}$ are

 \begin{eqnarray}
a_{11}&=&\left( \frac{\partial x}{\partial y'}\right)^2 + \left(\frac{\partial y}{\partial y'}\right)^2, ~ ~ a_{12}=-\left(\frac{\partial x}{\partial x'}\frac{\partial x}{\partial y'}+\frac{\partial y}{\partial x'}\frac{\partial y}{\partial y'}\right),\\
  a_{22}&=&\left(\frac{\partial x}{\partial x'}\right)^2+\left(\frac{\partial y}{\partial x'}\right)^2, ~ ~ a_{33}=\frac{\partial x}{\partial x'}\frac{\partial y}{\partial y'}-\frac{\partial x}{\partial y'}\frac{\partial y}{\partial x'}\label{eq:coefgeneral2}\end{eqnarray}
and finally the partial derivatives are as follows

\begin{eqnarray}
\frac{\partial x}{\partial x'} 
&=&
  \alpha {\Big[\alpha~ (x'^n-x_1^n)\Big]}^{\frac{1-n}{n}}~ x'^{n-1}\nonumber
\\
&+&\Big[ 1- \alpha^{\frac{1-n}{n}}~ (x'^n-x_1^n)^{\frac{1}{n}} \frac{{(x_2-x_0)}^{n-1}}{(x_2^n-x_1^n)}
\Big] \frac{\partial x_0}{\partial x'}\nonumber
\\
&+&  x_1^{n-1}~{\Big[\alpha~ (x'^n-x_1^n)\Big]}^{\frac{1-n}{n}}\Big[
\frac{(x_2-x_0)^n 
}{(x_2^n-x_1^n)^2}(x'^n-x_1^n) -\alpha
\Big] \frac{\partial x_1}{\partial x'}\nonumber
\\
&+&  \alpha^{\frac{1-n}{n}}~ (x'^n-x_1^n)^{\frac{1}{n}}{(x_2-x_0)}^{n-1}
\frac{(x_2^n-x_1^n) - (x_2-x_0) x_2^{n-1}
}{(x_2^n-x_1^n)^2}
\Big) \frac{\partial x_2}{\partial x'}
\end{eqnarray}

\begin{eqnarray}
\frac{\partial x}{\partial y'} &=&
\Big(1-
\alpha^{\frac{1-n}{n}}~ (x'^n-x_1^n)^{\frac{1}{n}} \frac{{(x_2-x_0)}^{n-1}(x_2^n-x_1^n)
}{(x_2^n-x_1^n)^2}
\Big)
\frac{\partial x_0}{\partial y'}\nonumber
\\
&+&~ x_1^{n-1}{[\alpha~ (x'^n-x_1^n)]}^{\frac{1-n}{n}} \Big((x'^n-x_1^n)
\frac{{(x_2-x_0)}^{n}
}{(x_2^n-x_1^n)^2}
- \alpha
\Big)
\frac{\partial x_1}{\partial y'}
\nonumber
\\
&+&{\alpha^{\frac{1-n}{n}}~ (x'^n-x_1^n)}^{\frac{1}{n}}{(x_2-x_0)}^{n-1}
~ \frac{x_0x_2^{n-1}-x_1^n
}{(x_2^n-x_1^n)^2}~
\frac{\partial x_2}{\partial y'}
\end{eqnarray}

\begin{eqnarray}
\frac{\partial y}{\partial x'} &=& \Big(1-
\beta^{\frac{1-n}{n}}~ (y'^n-y_1^n)^{\frac{1}{n}} \frac{{(y_2-y_0)}^{n-1}(y_2^n-y_1^n)
}{(y_2^n-y_1^n)^2}
\Big)
\frac{\partial y_0}{\partial x'}\nonumber
\\
&+&~ y_1^{n-1}{[\beta~ (y'^n-y_1^n)]}^{\frac{1-n}{n}} \Big((y'^n-y_1^n)
\frac{{(y_2-y_0)}^{n}
}{(y_2^n-y_1^n)^2}
- \beta
\Big)
\frac{\partial y_1}{\partial x'}
\nonumber
\\
&+&\beta^{\frac{1-n}{n}} ~ (y'^n-y_1^n)^{\frac{1}{n}}
 \frac{{(y_2-y_0)}^{n-1}(y_2^n-y_1^n) - (y_2-y_0)^n y_2^{n-1}
}{(y_2^n-y_1^n)^2}
\frac{\partial y_2}{\partial x'}
\end{eqnarray}

\begin{eqnarray}
\frac{\partial y}{\partial y'} 
&=&
  \beta {\Big[\beta~ (y'^n-y_1^n)\Big]}^{\frac{1-n}{n}}~ y'^{n-1}\nonumber
\\
&+&\Big[ 1- \beta^{\frac{1-n}{n}}~ (y'^n-y_1^n)^{\frac{1}{n}} \frac{{(y_2-y_0)}^{n-1}}{(y_2^n-y_1^n)}
\Big] \frac{\partial y_0}{\partial y'}\nonumber
\\
&+&  y_1^{n-1}~{\Big[\beta~ (y'^n-y_1^n)\Big]}^{\frac{1-n}{n}}\Big[
\frac{(y_2-y_0)^n 
}{(y_2^n-y_1^n)^2}(y'^n-y_1^n) -\beta
\Big] \frac{\partial y_1}{\partial y'}\nonumber
\\
&+&  \beta^{\frac{1-n}{n}}~ (y'^n-y_1^n)^{\frac{1}{n}}{(y_2-y_0)}^{n-1}
\frac{(y_2^n-y_1^n) - (y_2-y_0) y_2^{n-1}
}{(y_2^n-y_1^n)^2}
 \frac{\partial y_2}{\partial y'}
\label{eq:partialderivyy}
\end{eqnarray}

Now after having derived the general formulas for mimesis,  we supply the expressions of  the boundary points $(x_0,y_0),$ $(x_1,y_1)$, $(x_2,y_2),$   for some particular cases of regions $D_0,$  $D_1,$ $D_2$  needed for the numerical implementations. We finally turn our attention to the numerical simulations. The explicit  illustrations we have supplied to exemplify the work within this article have boundaries whose horizontal plane sections are parts of ellipses (sunflower-like petal, cross, circle) or lines (square, hexagram), see Fig. \ref{starpetal}.

 \subsection{Formulas for piecewise linear boundaries}\label{chap:linearboundaries}
In this section, we supply the explicit expressions of the coordinates $(x_0,y_0),$ $(x_1,y_1)$, $(x_2,y_2)$ of the boundary points used above, when the boundary of the star domain is piecewise linear.
 If a piece of the boundary of  a star domain is a line of the form $y=ax+b,$ then clearly the line through the origin and a point $(x',y')$ intersects this piece of boundary at $(x_i,y_i)=(\frac{bx'}{y'-ax'}, \frac{by'}{y'-ax'})$ and hence $\frac{\partial x_i}{\partial x'}=\frac{by'}{(y'-ax')^2}$, ~ $\frac{\partial x_i}{\partial y'}=-\frac{bx'}{(y'-ax')^2}$, ~ $\frac{\partial y_i}{\partial x'}=\frac{aby'}{(y'-ax')^2}$, ~ $\frac{\partial y_i}{\partial y'}=-\frac{abx'}{(y'-ax')^2}$.  Of course, in the case where this piece of boundary is a vertical segment with equation $x=c,$ then the above intersection is at  $(c, c\frac{y'}{x'}).$
 \subsection{Formulas for piecewise  elliptic boundaries}\label{chap:ellipses}
 We suppose here that a piece of at least one of our boundary curves is part of a nontrivial ellipse $\mathcal E_i$ with equation of the form $(x-a)^2/c_i^2+(y-b)^2/d_i^2$ where $(a,b)$ is our ventage point. Of course a line $(x(t),y(t))=(a,b) + t(x'-a,y'-b)$ passing through $(a,b)$ and a different point $(x',y')$  intersects $\mathcal E_i$ at two distinct points.
   For the construction, we need the point $(x_i,y_i)$ of that intersection which is nearer to $(x',y')$ in the sense that $x_i-a$ and $y_i-b$ have the same sign as $x'-a$ and $y'-b,$ respectively. We have \begin{eqnarray}x_i&=&a+\frac{x'-a}{\sqrt{(x'-a)^2/c_i^2+(y'-b)^2/d_i^2}}\\
  y_i&=&b+\frac{y'-b}{\sqrt{(x'-a)^2/c_i^2+(y'-b)^2/d_i^2}}
  \end{eqnarray}

So we have
\begin{eqnarray}
&\frac{\partial x_i}{\partial x'} = \frac{1}{\sqrt{\frac{(x'-a)^2}{c_i^2}+\frac{(y'-b)^2}{d_i^2}}}-\frac{(x'-a)^2}{\left( \frac{( x'-a)^2}{c_i^2}+\frac{(y'-b)^2}{d_i^2} \right)^{\frac{3}{2}}c_i^2}, \nonumber \\
&\frac{\partial x_i}{\partial y'} = -\frac{(x'-a)(y'-b)}{\left( \frac{( x'-a)^2}{c_i^2}+\frac{(y'-b)^2}{d_i^2} \right)^{\frac{3}{2}}d_i^2},
\end{eqnarray}

 and
\begin{eqnarray}
&\frac{\partial y_i}{\partial x'} =
-\frac{(x'-a)(y'-b)}{\left( \frac{(x'-a)^2}{c_i^2}+\frac{(y'-b)^2}{d_i^2}\right)^{\frac{3}{2}}c_i^2}, \nonumber \\
&\frac{\partial y_i}{\partial x'} = \frac{1}{\sqrt{\frac{(x'-a)^2}{c_i^2}+\frac{(y'-b)^2}{d_i^2}}}-\frac{(y'-b)^2}{\left( \frac{( x'-a)^2}{c_i^2}+\frac{(y'-b)^2}{d_i^2} \right)^{\frac{3}{2}}d_i^2}.
\end{eqnarray}

We can then apply formulas (\ref{eq:cylinders})-(\ref{eq:partialderivyy}) to build any generalized cloaks involving boundaries of elliptic types, where the center of the ellipse is the ventage point $(a,b)$. This will be used  in Fig.  \ref{starpetal}.

\section{Illustrative numerical results}

In this section, we apply our theory of non-linear transforms for cloaking and anamorphism of star shaped domains to illustrative cases. We start with the simple case of carpet cloaks for diffusive light. However, before we start looking at the design of cloaks, we would like to explain how we can rigorously model a diffusion problem, which is necessarily set in unbounded domain with a bounded computational domain.

\subsection{Perfectly Matched Layers through geometric transforms}\label{pmlheat}
To deal with the open problem, a judicious choice of coordinate
transformation allows the finite element modeling of the infinite exterior
domain \cite{nicolet94}.
Without loss of generality, we consider two disks $D(O,A)$ and
$D(O,B)$ of center $O = (0,0)$ and radii $A$ and $B>A$ strictly including
$\Omega$, we define a corona $C = D(O,B)\setminus \overline{D(O,A)}$, but the mapping can be applied to star domains.  Let
$(x,y)$ be a point in $\mathbb{R}^2\setminus \overline{D(O,A)}$ (the infinite outer
domain) and $(X,Y)$ be a point in $C$, the transformation is then given by:
$$
\cases{
x = f_1(X,Y) = X[A (B-A)] / [ R (B-R)] & \cr
y = f_2(X,Y) = Y[A (B-A)] / [ R (B-R)] & 
}
$$
where $R$ denotes the Euclidean norm $\sqrt{X^2 + Y^2}$. This
transformation may be viewed as a mapping of the finite corona $C$ with the non
orthogonal coordinate system $(X,Y)$ to the infinite domain with the cartesian
coordinate system $(x,y)$.

This way, the finite element discretization appears as a chained map from the reference space
to the transformed space and from the transformed space to the physical space.
It is also worth noting that taking Dirichlet or Neumann boundary conditions at a
finite distance (without geometric transformation) would change the physics of
the problem, by enforcing some values of the field at a finite distance. Importantly,
since we deal with a diffusion equation which is such that the field's amplitude is evanescent away from the
source, there is no need to consider a complex valued mapping like in \cite{nicolet08} for perfectly matched layers
(PMLs), which were originally introduced by Berenger \cite{berenger94} to absorb electromagnetic waves
propagating in all directions, without reflection. We finally point out that PMLs for elastodynamic waves require
some advanced mathematical formalism of scattered field theory \cite{diatta16}. We implement our PMLs for
diffusive light in fig. \ref{fig:carpet}, with a computational domain consisting of a square of sidelength $4$m, surrounded by
PMLs of width $0.5$m. The outer boundary of the computational domain has homogeneous Neumann data everywhere, except for the bottom, which has Dirichlet data.

\subsection{Anamorphism with carpet cloak for virtual source}
\begin{figure}[h]
\hspace{3cm}\mbox{}\scalebox{0.5}{
\includegraphics[width=20cm,angle=0]{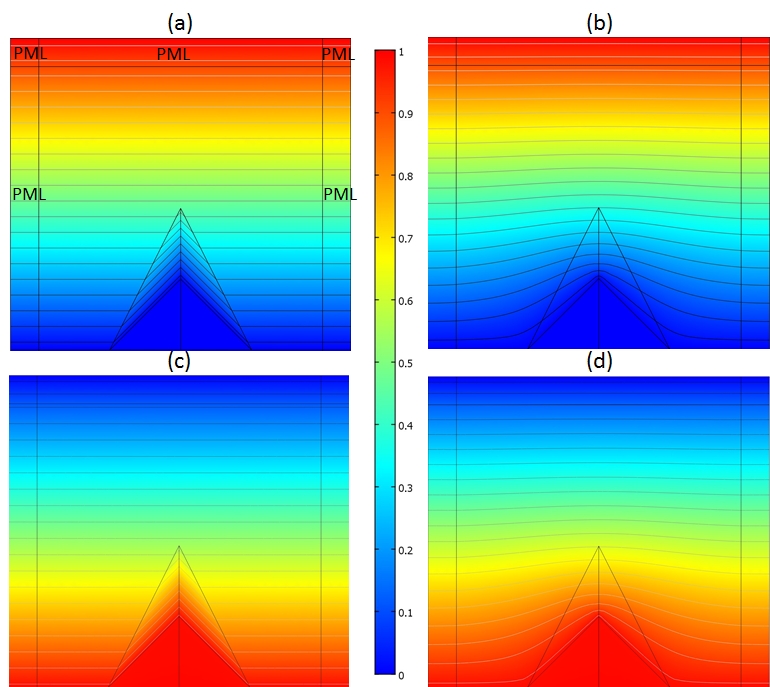}}
\caption{A simple star shape domain: A plane diffusive pseudo-wave propagating at angular frequency $\omega=\pi/50$ rad.s$^{-1}$ from top to bottom is incident upon an infinite conducting triangular object of height $1$m, and width $2$m with (a) and without (b) carpet of height $2$m, and width $2$m;
A plane diffusive wave is emitted at same frequency by the triangular object and propagates from bottom to top with (c) and without (d) the carpet. Note that (c) is a first example of anamorphism.}
\label{fig:carpet}
\end{figure}

We now consider a light diffusion carpet cloak of height $2$m, and width $2$m. The inner boundary of the carpet cloak is infinite conducting (transformed Neumann boundary conditions). We report these results in Fig. \ref{fig:carpet} for an angular finite frequency of $\omega=\pi/50$ rad.s$^{-1}$, which is appropriate for the carpet dimensions (one can easily scale down the size of carpet, and thus get high frequencies that would be in the visible spectrum as we deal with a linear problem).
Some homogeneous Dirichlet boundary conditions are set on the ground plane, the inner boundary of the carpet and the infinite conducting obstacle in panels (a), (b). However, in panels (c), (d), we set the source on the inner boundary of the carpet and the
object with a non-homogeneous Dirichlet data (with a normalized data of $1$), and one can see that it is virtually impossible for an external observer to deduce the shape of the source of diffusive light in (c), as it looks like it is a bare plane diffusive pseudo-wave. Let us emphasize that
we used our specially designed PMLs as detailed in section \ref{pmlheat}. 

\subsection{Anamorphism with circular cloaks: Squaring the circle}\label{chap:squarecircle}
\begin{figure}[h]
\hspace{3cm}\mbox{}\scalebox{0.4}{\includegraphics{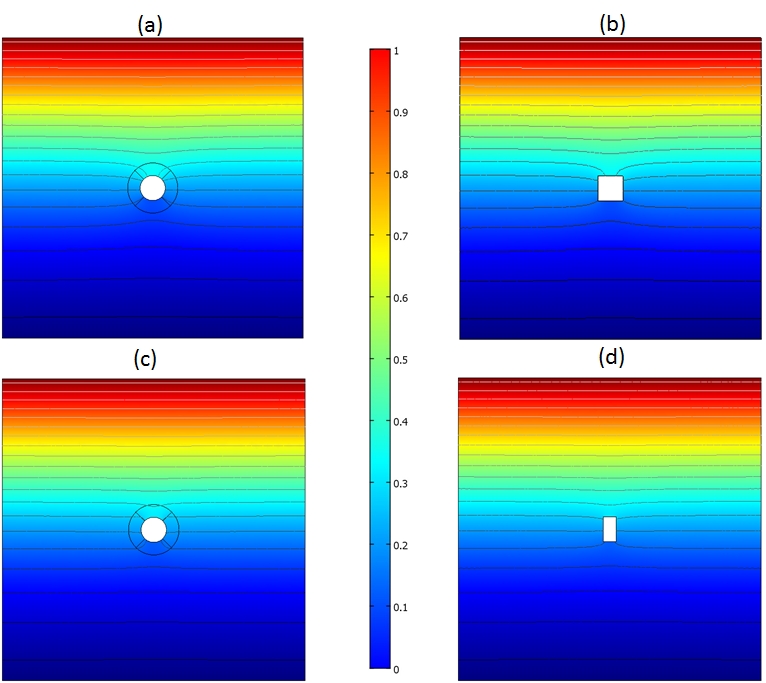}}
\caption{(Color online) Anamorphism for diffusive light: An infinite conducting circular object surrounded by a circular cloak (a) scatters diffusive light (a plane pseudo-wave) of frequency $\omega=\pi/10$ rad.s$^{-1}$ incident from top like an infinite conducting square object (b); An infinite conducting circular object surrounded by another circular cloak (c) scatters diffusive light like an infinite conducting rectangular object (d). See text for geometric data.}
\label{fig:mimetism1}
\end{figure}

\begin{figure}[h]
\hspace{3cm}\mbox{}\scalebox{0.43}{\includegraphics{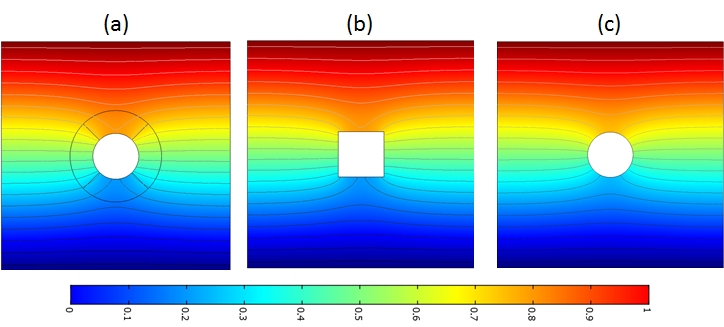}}
\caption{(Color online) Anamorphism for diffusive light: An infinite conducting circular object surrounded by a circular cloak (a) scatters diffusive light (a plane pseudo-wave) of frequency $\omega=\pi/10$ rad.s$^{-1}$ incident from top like an infinite conducting square object (b); An infinite conducting circular object (c) scatters diffusive light in a slightly different manner. See text for geometric data.}
\label{fig:mimetism1bis}
\end{figure}

In this section, we make a circle have the same (light diffusion) signature as an imaginary small square lying inside its enclosed region and sharing the same centre. In particular, its appearance to an external observer will look like that of a square. As above, the transformation will map the region enclosed between the small square  and the inner circle (cloaking surface) into the circular annulus bounded by the inside and outside circles, where the sides of the square are mapped to the inner circle and the outer one stays fixed point wise. To do so, we again use the diagonals of the square to part those regions into sectors. Indeed, the diagonals provide a natural triangulation by dividing the region enclosed inside the square into four sectors. The natural continuation of such a triangulation gives the needed one.
 In each sector,  we apply the same formulas  as above, where $(x_0,y_0)$ are obtained from the small square as in Section \ref{chap:linearboundaries} and for both $(x_1,y_1)$and $(x_2,y_2)$, we use the same formulas for ellipses in Section \ref{chap:ellipses}.
For the numerical simulation, we consider a pseudo-wave angular frequency $\omega=\pi/10$ rad.s$^{-1}$ in a computational square domain of sidelength $4$m with Dirichlet data on top and bottom, and Neumann data on left and right sides, respectively. We first consider a small square of side $L_0=0.4$m and two circles radii $R_1=0.2$m, $R_2=0.4$m centred at $(a,b)=(0,0)$, see Fig. \ref{fig:mimetism1} for a result of numerical simulations for the circular cloak (a) mimicking the square (b). We next consider in Fig. \ref{fig:mimetism1} a circular cloak with same boundaries (c) that now mimicks a rectangle of sidelengths $L_0=0.4$m and $l_0=0.2$m (d). The tiny, yet visible, differences in the fields in Fig. \ref{fig:mimetism1}, can be
made more pronounced by considering a smaller computational domain of sidelength $2$m. In Fig. \ref{fig:mimetism1bis} (a), we therefore consider the same circular cloak surrounding the circular scatterer as in Fig. \ref{fig:mimetism1} (a), which we compare to the scattering by the same square scatterer as in Fig. \ref{fig:mimetism1} (b), see Fig. \ref{fig:mimetism1bis} (b).
These two panels (a) and (b)  in Fig. \ref{fig:mimetism1bis} look exactly the same, however panel (c) of  Fig. \ref{fig:mimetism1bis}, which considers the scattering by an infinite conducting circular obstacle of radius $0.2$m, just like the one surrounded by the cloak in panel (a), is markedly different.   

\subsection{Star shaped and sunflower cloaks}
\begin{figure}[h]
\hspace{4cm}\mbox{}\scalebox{0.4}{\includegraphics{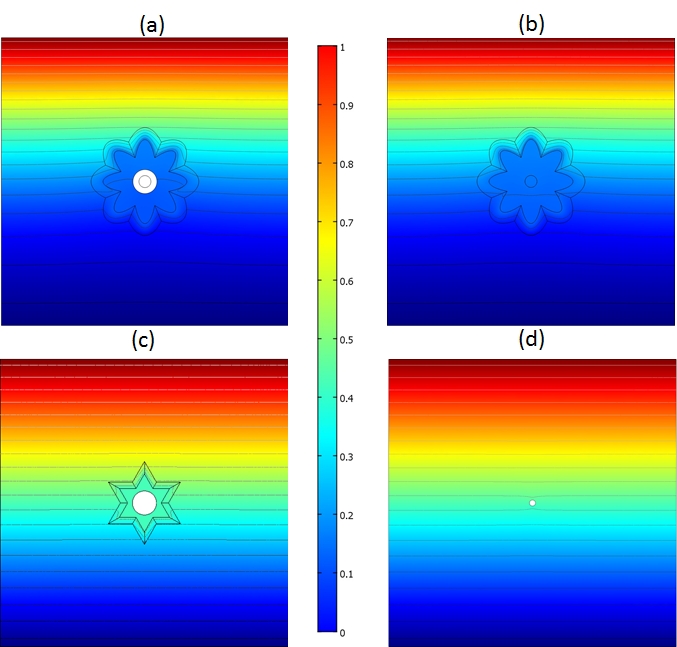}}
\caption{(Color online) A sunflower cloak (a) with inner and outer boundaries given by ellipses $x^2/0.7^2+y^2/0.2^2=1$ and $x^2/0.9^2+y^2/0.4^2=1$ (in units of meters) rotated by an angle $0$, $\pi/4$, $\pi/2$ and $3\pi/4$ rad surrounding an infinite conducting object of radius $r=0.2$m scatters a diffusive plane pseudo-wave of angular frequency $\omega=\pi/10$ rad.s$^{-1}$ incident from top like the same sunflower cloak now mimicking an infinite conducting object of radius $r=0.1$m (b); At an angular frequency $\omega=\pi/20$ rad.s$^{-1}$ a star shaped cloak hosting an infinite conducting object of radius $r=0.2$m (c) mimicks the scattering of an infinite conducting object of radius $r=0.05$m (d). One notes that the field is almost unperturbed outside the cloak and small (Neumann type) object in (c) and (d) and that the field is nearly constant in the sunfower and star shaped cloaks in (a), (b) and (c), but is not vanishing.} \label{starpetal}
\end{figure}

We now consider sunflower and star shaped light diffusion cloaks. We apply formulas (\ref{eq:cylinders})-(\ref{eq:partialderivyy}) to build the sunflower cloaks involving boundaries of elliptic types, where the center of each ellipse is a ventage point $(a,b)$ located on a radius starting from $(0,0)$. The sunflower cloaks have 8 petals, which are deduced from a rotation through an angle $\pi/4$. See Fig. \ref{starpetal} (a),(b) for further details on geometry and a numerical simulation of the diffusion light scattering of such cloaks at an angular frequency $\omega=\pi/10$ rad.s$^{-1}$. On the other hand, the design of star shaped cloaks requires an adapted triangulation of the corresponding region. That is, a triangulation that takes into account the singularities at the vertices of the boundary of the region. So that, each vertex belongs to the edge of some triangle. To the resulting triangles, one applies the same maps as in Sect. \ref{chap:transformation}.
See Fig. \ref{starpetal} (c), (d) for further details on geometry of star shaped cloak and result of numerical simulations at an angular frequency $\omega=\pi/20$ rad.s$^{-1}$, i.e. a quasi-static case.

\subsection{External cloaking for diffusive light}
External cloaking is rooted in a 1994 paper \cite{ext1} where a core-shell resonant system induced anomalous resonances that were later on use in the first external cloaking theory for small objects (dipoles) located in the close neighborhood of a cloak with a negatively refracting shell \cite{ext2}. However, this route did not work for finite size objects and finite frequencies \cite{ext3}. One possible way to extend external cloaking to finite obstacles and beyond the quasi-static limit is to make use of complemntary media, but this requires a priori knowledge of the obstacle to hide, as the cloak consists of its complementary medium \cite{ext4} (let us note in passing that complementary media for optical space cancellation were originally introduced in \cite{ext4b}). An alternative route to external cloaking makes use of a space folding that leads to spatially varying negtaively refracting shell \cite{ext5}. We propose here to adapt \cite{ext5} to the case of diffusion processes.    
Let us consider the 'space-folding' transform
\begin{eqnarray}
r=g(r')
\left\{
\begin{array}{lr}
r'r_c^2/r_s^2, ~ \rm{ for } ~r'\leq r_c \; ,
\\
r_s^2/r', ~ \rm{ for } ~ r_c\leq  r'\leq r_s \; ,
\\
r', ~ \rm{ for } ~ r' > r_s \; ,
\\
\end{array}
\right.
\label{yan1}
\end{eqnarray}
which leads to the transformed diffusivity and determinant of Jacobian
\begin{eqnarray}
\underline{\underline{D}'}=D{\bf T}^{-1}=
\left\{
\begin{array}{lr}
D{\bf I} \; \rm{ for } ~r'\leq r_c \; ,
\\
-D{\bf I} ~ \rm{ for } ~ r_c\leq r'\leq r_s \; ,
\\
D{\bf I} ~ \rm{ for } ~ r' > r_s \; ,
\\
\end{array}
\right. \\
\rm{ and \; det}({\bf J})=
\left\{
\begin{array}{lr}
r_s^4/r_c^4 \; \rm{ for } ~r'\leq r_c \; ,
\\
-r_s^4/r^4, ~ \rm{ for } ~ r_c\leq r'\leq r_s \; ,
\\
1, ~ \rm{ for } ~ r' > r_s \; ,
\\
\end{array}
\right.
\label{extcloak}
\end{eqnarray}
where ${\bf I}$ is the $2\times 2$ identity matrix.

Let us now detail the computation of the diffusion matrix (the reader versed in transformation optics techniques can skip this part, but we find it might be useful for newcomers in this field).
We start with the computation of the Jacobian matrix of the compound transformation
$(x,y,z)\rightarrow(r,\theta,z)\rightarrow(r',\theta',z')\rightarrow(x',y',z')$ which can be expressed as:
\begin{equation}
\begin{array}{ll}
{\bf J}_{xx'}
&={\bf J}_{xr}{\bf J}_{rr'}{\bf J}_{r'x'} \nonumber \\
&={\bf R}(\theta)\hbox{diag}(1,r,1)\hbox{diag}(g',1,1)\hbox{diag}(1,1/r',1){\bf R}(-\theta') \nonumber \\
&={\bf R}(\theta)\hbox{diag}(g',g/r',1){\bf R}(-\theta') \; ,
\end{array}
\end{equation}
where ${\bf J}_{xr}=\partial(x,y,z)/\partial(r,\theta,z)=\partial(r\cos\theta,r\sin\theta,z)/\partial(r,\theta,z)={\bf R}(\theta)\hbox{diag}(1,r,1)$,  ${\bf J}_{rr'}=\partial(r,\theta,z)/\partial(r',\theta',z')=\partial(g(r'),\theta,z)/\partial(r',\theta',z')$
and ${\bf J}_{r'x'}=\partial(r',\theta',z')/\partial(x',y',z')={\bf J}_{x'r'}^{-1}={[{\bf R}(\theta)\hbox{diag}(1,r',1)]}^{-1}=\hbox{diag}(1,1/r',1){\bf R}(-\theta')$ since the rotation matrix ${\bf R}(\theta')$ is such that ${\bf R}(\theta')^{-1}={\bf R}(-\theta')$.

\noindent We then compute the transformation matrix:
\begin{equation}
\begin{array}{ll}
{\bf T}^{-1}&={\bf J}_{xx'}^{-1}{\bf J}_{xx'}^{-T}\hbox{det}({\bf J}_{xx'}) \nonumber \\
&={\bf R}(-\theta')^{-1}\hbox{diag}(1/g',r'/g,1){\bf R}(\theta)^{-1} \nonumber \\
&{\bf R}(\theta)^{-T}\hbox{diag}(1/g',r'/g,1){\bf R}(-\theta')^{-T} (gg')/r' \nonumber \\
&={\bf R}(\theta')\hbox{diag}(g/(g'r'),(g'r')/g,(gg')/r'){\bf R}(\theta')^T \; ,
\end{array}
\label{bleu1}
\end{equation}
where we have used the fact that the rotation matrix satisfies ${\bf R}(\theta)^{-1}={\bf R}(\theta)^T$
and $\theta'=\theta$.

\begin{figure}[h]
\hspace{2cm}\mbox{}\scalebox{1.0}{
\includegraphics[width=14cm,angle=0]{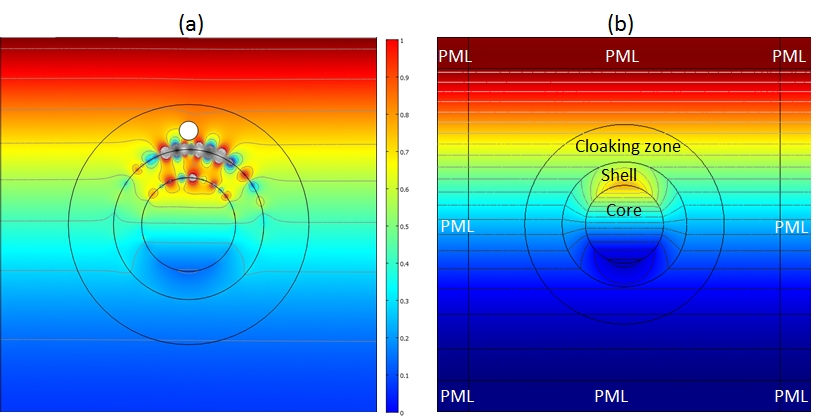}}
\caption{External light diffusion cloak with spatially varying sign-shifting isotropic parameters as in equation (\ref{extcloak}) reducing the scattering by an infinite conducting object located
outside the cloak (a) via anomalous resonances at the inner and outer boundaries of the cloak, which is otherwise undetectable (b). Note that the outer black circle denotes the (fictitious) boundary of cloaking region.}
\label{fig:external}
\end{figure}

\noindent We now note that from (\ref{yan1}) one has:

\begin{eqnarray}
&\frac{g(r')}{g'(r')r'}=
\left\{
\begin{array}{lr}
1, ~ \rm{ for } ~r'\leq r_c \; ,
\\
-1, ~ \rm{ for } ~ r_c\leq  r'\leq r_s \; ,
\\
1, ~ \rm{ for } ~ r' > r_s \; ,
\\
\end{array}
\right.
\; \nonumber \\
&\rm{ and }
\;
\frac{g(r')g'(r')}{r'}=
\left\{
\begin{array}{lr}
r_s^4/r_c^4, ~ \rm{ for } ~r'\leq r_c \; ,
\\
-r'^4/r_s^4, ~ \rm{ for } ~ r_c\leq  r'\leq r_s \; ,
\\
1, ~ \rm{ for } ~ r' > r_s \; .
\\
\end{array}
\right.
\label{yan2new}
\end{eqnarray}
Results of finite element computations are shown in Fig. \ref{fig:external} for radii $r_c=0.5$m, $r_s=0.8$m and a finite angular frequency of $\omega=\pi/10$ rad.s$^{-1}$ in panel (a) and $\pi/2$ rad.s$^{-1}$ in panel (b), where we note that we used our specially designed PMLs as detailed in section \ref{pmlheat}. 

Such a design of external cloak for light of mass diffusion might seem far-fetched, but we note that propagation of photon density waves in random amplifying media studied in \cite{anantha15} with DPDW makes a theory consistent with physics of negatively refracting media \cite{anantha15,anantha16}, notably versus space folding and lensing \cite{kadic11} that is aking to apparent negative thermal diffusivity in \cite{huang}.

\section{A non-linear transform for form invariant Fick's equation of mass diffusion}
Let us now assume that $\nu=1$ and $\mu_a=0$ in (\ref{govpressure}), this leads us to Fick's equation, where
$D$ is a chemical diffusivity in units of $m^3.s^{-1}$ and $\Phi$ a concentration. In \cite{guenneau13}, where
the transformed Fick's equation counterpart of (\ref{transfpressure}) with $\nu=1$ and $\mu_a=0$
was studied in the time-domain, it was noted that the parameter
$\nu'=T_{zz}\nu=\rm{det}({\bf J})$ has no physical meaning for $\nu'\ne 1$.

In order to overcome this pitfall, we now consider a geometric transform for cloaking which is such that \rm{det}({\bf J}) is a constant.
This can be done assuming that $n=2$ in (\ref{eqnonlinear}), and further considering some boundaries parameterized by an azimuthal angle $0<\theta\leq 2\pi$:
\begin{equation}
\begin{array}{ll}
&\rho'(\rho,\theta)=\sqrt{\frac{r_2^2(\theta)-r_1^2(\theta)}{{(r_2(\theta)-r_0(\theta))}^2}{(\rho-r_0(\theta))}^2+r_1^2(\theta)} \; , \nonumber \\
&r_0(\theta)\leq r\leq r_2(\theta) \; , \; \theta'=\theta \; , \; z'=z \; ,
\end{array}
\label{magictransfo}
\end{equation}
whose Jacobian is such that $\rm{det}({\bf J})(\theta)=1/\sqrt{\frac{r_2^2(\theta)-r_1^2(\theta)}{{(r_2(\theta)-r_0(\theta))}^2}}$.

If we further consider some radially symmetric (circular) boundaries, we can see that $\rm{det}({\bf J})$ becomes a constant.
Interestingly, if we take $r_0=0$, we end up with the nonlinear counterpart of Pendry's formula
\begin{equation}
r'=\sqrt{(1-r_1^2/r_2^2)r^2+r_1^2} \; , \; 0\leq r\leq r_2 \; , \; \theta'=\theta \; , \; z'=z \; ,
\label{magictransfo}
\end{equation}
which has Jacobian
\begin{equation}
{\bf J}_{xx'}=\frac{1}{\sqrt{1-r_1^2/r_2^2}}{\bf R}(\theta){\rm Diag}\left(\frac{r'}{\sqrt{r'-r_1^2}},\frac{\sqrt{r'-r_1^2}}{r'}\right){\bf R}(-\theta') \; ,
\label{jac1}
\end{equation}
which is such that $\rm{det}({\bf J}_{xx'})={(1-r_1^2/r_2^2)}^{-1/2}$.

Provided that $\rm{det}({\bf J}_{xx'})$ is constant (what requires circular boundaries), it is possible to recast  (\ref{transfpressure}) as
\begin{equation}
-\nabla\cdot \left(\underline{\underline{D''}}({\bf x'},t) \nabla \Phi({\bf x'},t)\right) - \frac{i\omega}{\nu({\bf x'})}\Phi({\bf x'},t) + \mu_{a}({\bf x'}) \Phi({\bf x'},t)  = S({\bf x'},t) \Phi({\bf x'},t)
\; , \label{transfpressure1bis}
\end{equation}
with $\underline{\underline{D''}}({\bf x'})=D\rm{det}({\bf J}^{-1}){\bf  T}_T^{-1}=D{\bf J}_{xx'}^{-1}{\bf J}_{xx'}^{-T}$.
When $\mu_a=0$ and $\nu=1$, Fick's equation is form invariant under (\ref{magictransfo}).

We note that (\ref{magictransfo}) also solves the issue of reduced parameters that brought a slight discrepancy in Fourier's equation in \cite{guenneau12}, which were used for a layered thermal cloak via
homogenization. Unfortunately, the form invariant Fick's equation under (\ref{magictransfo}) holds only for circular boundaries, and breaks down in the spherical case.

Let us now detail the computation of the diffusion matrix. We start with the computation of the Jacobian matrix of the compound transformation
$(x,y)\rightarrow(r,\theta)\rightarrow(r',\theta')\rightarrow(x',y')$ which can be expressed as
\footnote{Noting that $r'=\sqrt{\alpha r^2+r_1^2}$ hence $r=\alpha^{-1/2}\sqrt{{r'}^2-r_1^2}$, we can see that (\ref{todo}), simplifies into ${\bf J}_{xx'}=\alpha^{-1/2}{\bf R}(\theta)\hbox{diag}(r'/\sqrt{{r'}^2-r_1^2},\sqrt{{r'}^2-r_1^2}/r'){\bf R}(-\theta')$.}:
\begin{equation}
\begin{array}{ll}
{\bf J}_{xx'}={\bf J}_{xr}{\bf J}_{rr'}{\bf J}_{r'x'}
&={\bf R}(\theta)\hbox{diag}(1,r)\hbox{diag}(r'/(r\alpha),1)\hbox{diag}(1,1/r'){\bf R}(-\theta') \nonumber \\
&={\bf R}(\theta)\hbox{diag}(r'/(r\alpha),r/r'){\bf R}(-\theta') \; ,
\end{array}
\label{todo}
\end{equation}
where ${\bf J}_{xr}=\partial(x,y)/\partial(r,\theta)$,  ${\bf J}_{rr'}=\partial(r,\theta)/\partial(r',\theta')$
and ${\bf J}_{r'x'}=\partial(r',\theta')/\partial(x',y')$; Moreover, $\alpha=(r_2^2-r_1^2)/r_2^2$ and we note
that $\hbox{det}({\bf J}_{xx'})=\alpha^{-1}$ as the rotation matrices ${\bf R}(\theta)$ and
${\bf R}(\theta')$ are unimodular.

\noindent We then compute the transformation matrix:
\begin{equation}
\begin{array}{ll}
{\bf T}^{-1}&={\bf J}_{xx'}^{-1}{\bf J}_{xx'}^{-T}\hbox{det}({\bf J}_{xx'}) \nonumber \\
&={\bf R}(-\theta')^{-1}\hbox{diag}((r\alpha)/r',r'/r){\bf R}(\theta)^{-1} \nonumber \\
&{\bf R}(\theta)^{-T}\hbox{diag}((r\alpha)/r',r'/r){\bf R}(-\theta')^{-T} \alpha^{-1}\nonumber \\
&={\bf R}(\theta')\hbox{diag}((\alpha r^2)/r'^2,r'^2/(r^2\alpha)){\bf R}(\theta')^T \; ,
\end{array}
\label{bleu}
\end{equation}
where we have used the fact that the rotation matrix satisfies ${\bf R}(\theta)^{-1}={\bf R}(\theta)^T={\bf R}(-\theta)$
and $\theta'=\theta$.

\noindent We now observe that from (\ref{magictransfo}) $r^2=(r'^2-r_1^2)/\alpha$, hence the transformed diffusivity inside
the circular coating of the cloak can be expressed as
\begin{equation}
\underline{\underline{D}'}=D{\bf T}^{-1}
=D{\bf R}(\theta')\hbox{diag}(D'_{r'},D'_{\theta'}){\bf R}(\theta')^T \; ,
\end{equation}
where the eigenvalues of the diagonal matrix (principal values of conductivity) are from (\ref{bleu})
\begin{equation}
\begin{array}{lll}
D'_{r'} &=\displaystyle{\frac{r'^2-r_1^2}{r'^2}} \; ,
& D'_{\theta'}=\displaystyle{{\frac{r'^2}{r'^2-r_1^2}}} \; ,
\end{array}
\label{rhort1a}
\end{equation}
with $r_1$ and $r_2$ the inner and the outer radii of the cloak.

In order to eliminate $T_{zz}=\rm{det}({\bf J})$ in the factor of $i\omega\Phi$ as in (\ref{transfpressure1bis}), we multiply the eigenvalues by $\rm{det}({\bf J})$ and find
\begin{equation}
\begin{array}{lll}
D'_{r'} &=\displaystyle{\sqrt{1-r_1^2/r_2^2}\frac{r'^2-r_1^2}{r'^2}} \; ,
& D'_{\theta'}=\displaystyle{\sqrt{1-r_1^2/r_2^2}{\frac{r'^2}{r'^2-r_1^2}}} \; ,
\end{array}
\label{rhort1aa}
\end{equation}

This should be compared with eigenvalues of diffusivity and determinant of Jacobian obtained from
Pendry's transform when $n=1$ \footnote{Noting that for Pendry's transform $r'=\beta r+r_1$, with $\beta=(r_2-r_1)/r_2$ hence $r=\beta^{-1}({r'}-r_1)$, we can see that ${\bf J}_{xx'}
={\bf R}(\theta)\hbox{diag}(1,r)\hbox{diag}(1/\beta,1)\hbox{diag}(1,1/r'){\bf R}(-\theta')
={\bf R}(\theta)\hbox{diag}(1/\beta),r/r'){\bf R}(-\theta')$, and det(${\bf J}_{xx'}$)$=r/(\beta r')=(r'-r_1)/(\beta^2 r')$ and ${\bf T}^{-1}=
{\bf R}(-\theta')^{-1}\hbox{diag}(\beta^2,{(r'/r)}^2){\bf R}(-\theta')^{-T} (r/(\beta r'))={\bf R}(-\theta')^{-1}\hbox{diag}((r\beta)/r',r'/(r\beta)){\bf R}(\theta)^{-1}={\bf R}(-\theta')^{-1}\hbox{diag}((r'-r_1)/r',r'/(r'-r_1)){\bf R}(\theta)^{-1}$.}:
\begin{equation}
\begin{array}{lll}
D'_{r'} &=\displaystyle{\frac{r'-r_1}{r'}} \; ,
& D'_{\theta'}=\displaystyle{{\frac{r'}{r'-r_1}}} \; ,
\rm{det}{\bf J}= \displaystyle{\frac{r'-r_1}{r'}{\left(\frac{r_2}{r_2-r_1}\right)}^2}
\end{array}
\label{rhort1aaa}
\end{equation}
Note that  $(\rm{det}{\bf J})$ vanishes for $r' = r_1$, in which case the frequency dependence is discarded. Physically, replacing the frequency by a time derivative, this means it should take an infinite amount of time for mass to cross the inner boundary of the cloak. This led to counterintuitive results for the transformed heat equation in \cite{guenneau12}, where it was observed that heat inside the invisibility region is constant at each time step, and similar results hold for the transformed Fick's equation in \cite{guenneau13}. We would like to investigate what happens inside this invisibility region with the nonlinear transform that avoids the vanishing determinant. 

\section{Non-linear transform for form invariant diffusion equations in time domain}
In this section, we draw interesting consequences on the application of non-linear transform (\ref{magictransfo}) to various diffusion equations in the time domain.

\subsection{Form invariant Fick, Fourier and Schr\" odinger equations}
Provided that $\rm{det}({\bf J})$ is constant (what requires circular boundaries), the transformed light diffusion equation in time domain writes as
\begin{equation}
-\nabla\cdot \left(\underline{\underline{D''}}({\bf x'}) \nabla \Phi({\bf x'})\right) + \frac{\partial \Phi({\bf x'})}{\nu({\bf x'})\partial t} + \mu_{a}({\bf x'}) \Phi({\bf x'})  = S({\bf x'}) \Phi({\bf x'})
\; , \label{transfpressure1}
\end{equation}
with $\underline{\underline{D''}}({\bf x'})=D\rm{det}({\bf J}^{-1}){\bf  T}_T^{-1}=D{\bf J}_{xx'}^{-1}{\bf J}_{xx'}^{-T}$.
When $\mu_a=0$ and $\nu=1$, the time dependent Fick's equation is form invariant under (\ref{magictransfo}), which was not the case in \cite{guenneau13}.

The transformed Fick's equation in \cite{guenneau13} has served as an inspiration for the work on light diffusion cloaks by Wegener's group, and using our geometric transform (\ref{magictransfo}), it should be now possible to achieve well behaved cloaks in the transient domain, since there is no longer the issue of the heterogeneous determinant as a factor of the time derivative. Moreover, Fourier's equation for heat also takes a nicer form, when using our transform:
 \begin{equation}
-\nabla\cdot \left(\underline{\underline{D''}}({\bf x'}) \nabla \Phi({\bf x'})\right) + \rho({\bf x'})c_p({\bf x'})\frac{\partial T({\bf x'})}{\partial t} = S({\bf x'}) \Phi({\bf x'})
\; , \label{heat}
\end{equation}
as now there is no more issue with vanishing values of product of density by specific heat at the inner boundary of the cloak. Therefore, the thermodynamic cloak design experimentally validated in \cite{schittny13} might be further improved. 

Finally, one should note that there is a strong analogy between (\ref{transfpressure1}) and the time-dependent Schr\" odinger equation in the context of quantum mechanics. Some time-independent quantum cloaks for matter waves have been proposed by the groups of Zhang and Greenleaf using coordinate transforms \cite{matter_prl2008}, but
they suffered from an inherent singularity due to the vanishing determinant on the inner boundary of the cloak, which is overcome with the non-linear transform using the
transformed Schr\" odinger equation
\begin{equation}
\hbox{det}({\bf J}) \left ( i\hbar\frac{\partial\Psi}{\partial t} \right)=
-\frac{\hbar^2}{2} \nabla\cdot\left({\bf J}^{-1}m^{-1}({\bf x}') {\bf J}^{-T}\hbox{det}({\bf J})\nabla \Psi\right)
+ \hbox{det}({\bf J})V\Psi \; ,
\label{transcat}
\end{equation}
where $\hbox{det}({\bf J})$ can now be safely removed from the equation, which has not been studied thus far in the cloaking literature. Here, what plays the role of the transformed conductivity in
(\ref{transfpressure1}) is $ \underline{\underline{m'}}^{-1}={\bf J}^{-1}m^{-1}({\bf x}') {\bf J}^{-T}\hbox{det}({\bf J})$,
which is a transformed (anisotropic heterogeneous) mass.
Note that when $\hbox{det}({\bf J})$ is not a constant, $\hbox{det}({\bf J})V$
is a transformed (heterogeneous) potential. Moreover, $\hbox{det}({\bf J})$
in the left-hand side is a time-scaling parameter,
$\hbar$ is the Planck constant and $\Psi$ is the wave function. In order to facilitate the physical interpretation
of (\ref{transcat}), one can now divide throughout by $\hbox{det}({\bf J})$, since it is a constant thanks
to  (\ref{magictransfo}).

\subsection{Homogenization route to anisotropic diffusivity}
We have considerably improved the cloak's design, since we do not have to handle the heterogeneous determinant
like in \cite{guenneau12,guenneau13,petiteau14}. Nonetheless, we still face the obstacle of an anisotropic diffusivity
inside the cloak. This problem can be handled using techniques of homogenization see e.g. chapter 1 in \cite{bensoussan78,boutin09,craster12} for a detailed derivation in the case of the conductivity and wave equations. The book of Milton on
the theory of composites also contains many results on homogenization of layered media, that can be of interest to the reader \cite{milton02}.
Let us consider that the field is solution of the following DPDW equation
(outside the source)
\begin{equation}
\begin{array}{ll}
&-\displaystyle{\nabla\cdot \left(D({\bf x},{\bf x}/\eta) \nabla \Phi_\eta({\bf x})\right) + \frac{\partial \Phi_\eta({\bf x})}{\nu({\bf x},{\bf x}/\eta)\partial t}} \nonumber\\
&+ \mu_{a}({\bf x},{\bf x}/\eta) \Phi_\eta({\bf x})  = S({\bf x}) \Phi_\eta({\bf x})
\end{array}
\; , \label{transfpressure1}
\label{acoustic-hom1}
\end{equation}
inside the heterogeneous isotropic cloak $\Omega_f$, where
$D({\bf x},{\bf y})$, $\mu_{a}({\bf x},{\bf y})$ and $\nu({\bf x},{\bf y})$ are piecewise continuous functions of period $1$ in ${\bf y}$.
When $0<\eta\ll 1$, the field $\Phi_\eta({\bf x})$ oscillates rapidly in the structured cloak $\Omega_f$, with oscillations on the
order of $\eta$. To filter these variations, we consider an asymptotic expansion of $\Phi_\eta$
solution of (\ref{acoustic-hom1}) in terms of a macroscopic (or slow) variable ${\bf x}=(r,\theta)$ and a
microscopic (or fast) variable ${\bf
x}_\eta=(\frac{r}{\eta},\theta)$.

\noindent The homogenization of DPDW equation can be derived by considering
the ansatz
\begin{equation}
\Phi_\eta ({\bf x}) =
\sum_{i=0}^\infty \eta^i \Phi^{(i)}({\bf x},{\bf x}_\eta)
\; ,
\label{ansatz}
\end{equation}
where for every ${\bf x}\in\Omega_f,\; u^{(i)}({\bf x},\cdot)$
is 1-periodic along $r$. Note that we evenly divide $\Omega_f$
($R_1\leq r \leq R_2$, $0\leq\theta<2\pi$) into a large number of
thin curved layers of radial length $(R_2-R_1)/\eta$, but the
time variable $t$ in (\ref{acoustic-hom1}) is not rescaled.

\noindent Rescaling the differential operator in Eq. (\ref{acoustic-hom1})
accordingly as $\nabla=\nabla_{\bf x}+{\bf
e}_r\displaystyle{\frac{1}{\eta}\frac{\partial}{\partial r}}$, and
collecting the terms sitting in front of the same powers of $\eta$, we obtain
\begin{equation}
\begin{array}{ll}
&-\displaystyle{\nabla\cdot \left(\underline{\underline{D_{hom}}}({\bf x}) \nabla \Phi_0({\bf x})\right) + \left <  \frac{1}{\nu} \right > ({\bf x}) \frac{\partial \Phi_0({\bf x})}{\partial t}} \nonumber \\
&+ <\mu_{a}>({\bf x}) \Phi_0({\bf x})  = S({\bf x}) \Phi_0({\bf x})
\end{array}
\label{acoustic-hom2}
\end{equation}
which is the homogenized DPDW equation
where $\Phi_0$ is the leading order term in the ansatz (\ref{ansatz}).
Here, $<f>({\bf x})=\int_Y f({\bf x},{\bf y}) \, d{\bf y}$ is the arithmetic mean over a unit cell $Y$ along the radial axis
and $\underline{\underline{D_{hom}}}$ is a homogenized rank-2 diagonal tensor,
which has the physical dimensions of a homogenized anisotropic diffusivity
$\underline{\underline{D_{hom}}}={\rm Diag}(D_r,D_\theta)$ given by
\begin{equation}
\underline{\underline{D_{hom}}}={\rm
Diag}({<D^{-1}>}^{-1},<D>)
\; .
\label{parameter2}
\end{equation}

\noindent We note that if the cloak consists of an alternation of two
homogeneous isotropic layers of thicknesses $d_A$ and $d_B$ and
diffusivities $D_A$, $D_B$, light speed $\nu_A$, $\nu_B$
and absorption coefficients $\mu_{a,A}$, $\mu_{a,B}$, we obtain the following
effective parameters
\begin{equation}
\begin{array}{lll}
&\displaystyle{\frac{1}{D_r}}=\displaystyle{\frac{1}{1+\eta}\left(\frac{1}{D_A}+\frac{\eta}{D_B}\right)} \; ,
\; D_\theta=\displaystyle{\frac{D_A+\eta D_B}{1+\eta}} \; , \; \nonumber \\
&\displaystyle{ \left <  \frac{1}{\nu} \right > } = \displaystyle{\frac{\nu_A^{-1}+\eta \nu_B^{-1}}{1+\eta}} \; ,
\; <\mu_{a}>= \displaystyle{\frac{\mu_{a,A}+\eta \mu_{a,B}}{1+\eta}}
\nonumber
\end{array}
\end{equation}
which are respectively described by one harmonic and three arithmetic means,
where $\eta=d_B/d_A$ is the ratio of thicknesses for layers $A$ and
$B$ and $d_A+d_B=1$.

\subsection{Time dependence of field inside the invisibility region}

\begin{figure}[h]
\hspace{2cm}\mbox{}\scalebox{0.4}{\includegraphics{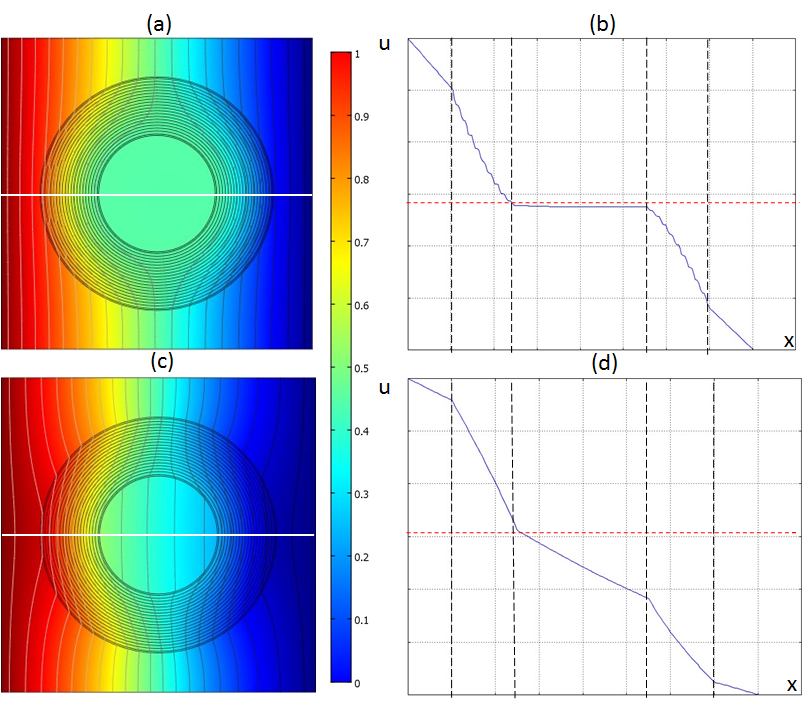}}
\caption{(Color online) Distribution of mass diffusion in a multilayered invisibility cloak with 2D plot (a), profile along horizontal axis (b)
compared against a low diffusivity homogeneous medium (c,d).  Time step=40ms.}
\label{fig:mimetism7b}
\end{figure}

\begin{figure}[h]
\hspace{2cm}\mbox{}\scalebox{0.4}{\includegraphics{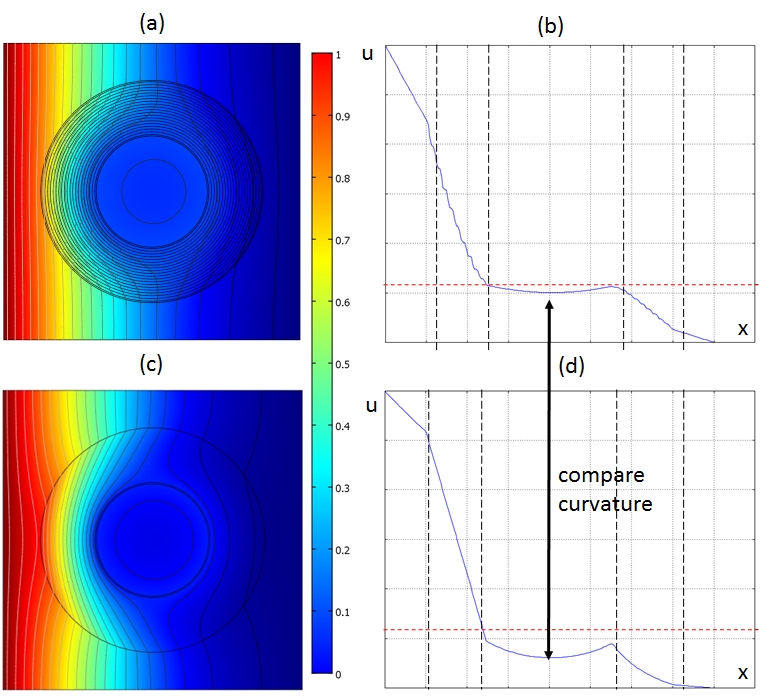}}
\caption{(Color online) Distribution of mass diffusion in a multilayered invisibility cloak with 2D plot (a), profile along horizontal axis (b)
compared against a low diffusivity homogeneous medium (c,d).  Time step=2ms.Note the more pronounced curvature in (d) than (b).}
\label{fig:mimetism7c}
\end{figure}

\begin{figure}[h]
\hspace{2cm}\mbox{}\scalebox{0.4}{\includegraphics{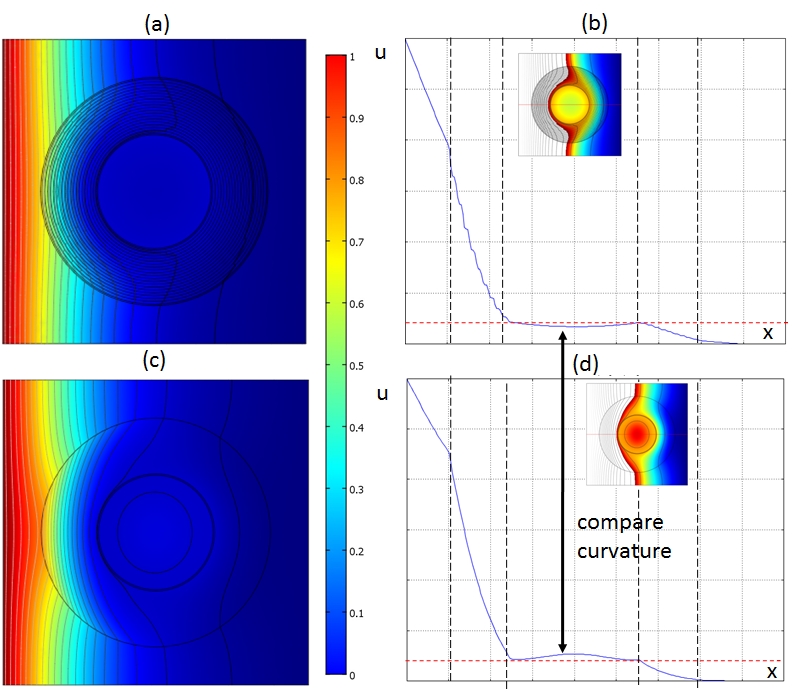}}
\caption{(Color online) Distribution of mass diffusion in a multilayered invisibility cloak with 2D plot (a), profile along horizontal axis (b)
compared against a low diffusivity homogeneous medium (c,d).  Time step=5ms. Note the opposite curvature in (d) and (b).}
\label{fig:mimetism7d}
\end{figure}

\begin{figure}[h]
\hspace{2cm}\mbox{}\scalebox{0.4}{\includegraphics{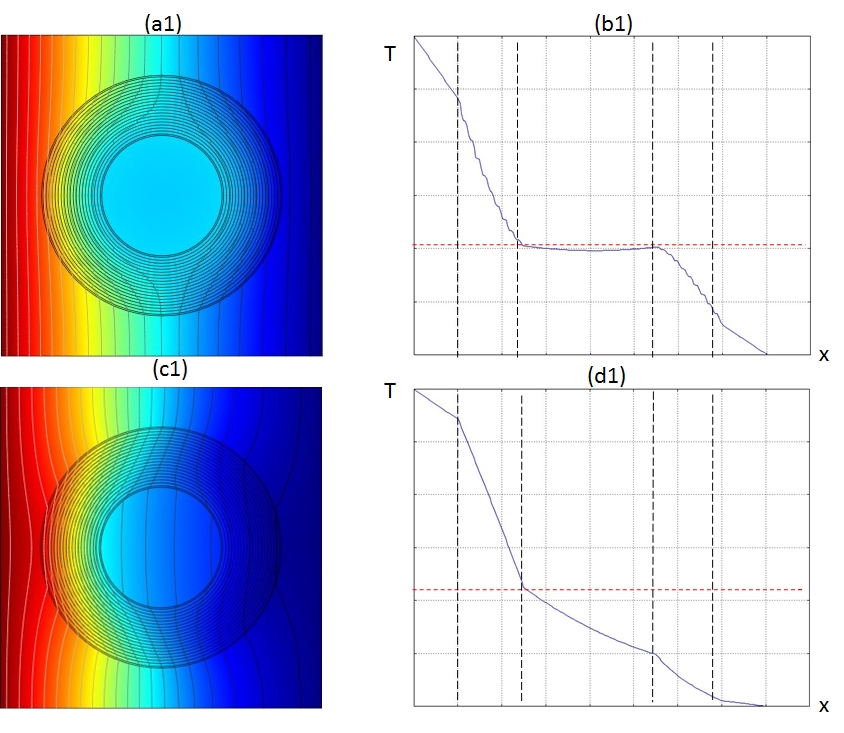}}
\caption{(Color online) Distribution of mass diffusion in a multilayered invisibility cloak with 2D plot (a), profile along horizontal axis (b)
compared against a low diffusivity homogeneous medium (c,d). Time step=10ms.}
\label{fig:mimetism7a}
\end{figure}

We consider a multilayered cloak with radii $r_1=150$ mm and $r_2=300$mm, with $20$ layers of diffusivity varying between $0.25$ and $1680$ (in units of $10^{-5} m^{2}.s^{-1}$)
and compare its diffusion against a shell with same radii and a thin inner layer of diffusivity $0.25$ (graphene could be used) and a thick outer layer of diffusivity $1680$ (still in units of $10^{-5} m^{2}.s^{-1}$). We show in Figures \ref{fig:mimetism7b}-\ref{fig:mimetism7a} the spatial distribution of mass at representative time steps (this is similar for light, heat etc., except that one has to apply a scaling factor on time due to the very diffusion speed for chemical species, light, heat etc.). We start with the case of long times, when we reach the stationary regime in Figure \ref{fig:mimetism7b}, where we note in panels (a) and (b) that the field is uniform inside the invisibility region of the cloak, unlike for the two-layer shell. We plot the profile of mass concentration in panels (b) and (d) along the x-axis highlighted by the white segment in panels (a) and (b). However, when we look at short times, we can see quite interesting antagonistic behaviours in the cloak and the shell, see Figures \ref{fig:mimetism7d}-\ref{fig:mimetism7c}: field increases with time almost uniformly in the invisibility region in panels (a) and (b), whereas it varies non monotonically in panels (b) and (c), and eventually reaches a uniform gradient. Let us explain this phenomenon.

If we assume that the field (mass, or photon density or temperature) on the
inner disc $D$ is zero, then the weak
maximum principle ensures us that
the temperature is zero everywhere
inside the disc $D$. However, if the
temperature on the boundary
$\partial D$ of $D$ 
this theorem only states that
$\max_{D} u=\max_{\partial D}u$.
In order to demonstrate that the field
inside the disc is a constant for any given
time $t$, we need to apply the
mean value property, which states that
\cite{max1}:

\noindent{\bf Theorem (Nirenberg):}
{\it Let $u\in C_1^2(D\times[0,T))$ (i.e.
continuous with first and second space derivatives continuous
and first time derivative continuous) solve the heat
equation (\ref{heat}) then:
\begin{equation}
u(x,y,t)= \frac{\lambda}{4}\int_{E_\lambda(x,y,t)} u(v,w,s) \frac{(x-v)^2+(y-w)^2}{{(t-s)}^2} \, dvdwds \; ,
\label{max1}
\end{equation}
for each $E_\lambda(x,y,t)\subset D$. Here, $E_\lambda(x,y,t)$ is the so-called heat ball
which is a super-level set of the fundamental solution of the heat equation
defined as
$E_\lambda(x,y,t)=\{(v,w,s)\in \mathbb{R}^2\times [0,\infty) \mid s\leq t \; , \;
\Phi(x-v,y-w,t-s)\geq \lambda\}$
and
$\Phi(x,y,t)={(4t\pi)}^{-1}\exp(-(x^2+y^2)/t)$.}

\noindent As a corollary of this theorem, if D is connected and there
exists $(x_0,y_0,t_0) \in D\times[0,T)$ such that
$u(x_0,y_0,t_0)=\max_{\overline{D\times[0,T)}}u$,
then by picking $r$ small enough so that
$E_\lambda((x_0,y_0),t_0)\subset D\times[0,T)$,
and using the mean value property, we
conclude that $u$ is constant inside
$E_\lambda(x_0,y_0,t_0)$.

\noindent Next for any $(u_0,v_0,s_0)\in D\times[0,T)$
such that the line segment connecting $(x_0,y_0)$ to $(v_0,w_0)$ is in $D$, we can show that
$u(v_0,w_0,s_0)=u(x_0,y_0,t_0)$
whenever
$s_0<t_0$
by covering the line segment connecting
$(v_0,w_0,s_0)$
and
$(x_0,y_0,t_0)$
with the heat balls.

\noindent Finally, since $D$ is connected,
any $(v_0,w_0)$ can be connected
from $(x_0,y_0)$ via
finitely many line segments.
And therefore we have
$u(v,w,s)=u(x_0,y_0,t_0)$
for all $(v,w)\in D$, $s<t_0$.

\noindent We have therefore shown that if the
invisibility region $D$ is connected, which is
clearly the case for a disc, but also for
the invisibility region shown in Figures \ref{fig:mimetism7b}-\ref{fig:mimetism7a},
and if there
exists $(x_0,y_0)\in D$ such that $u(x_0,y_0)=\max_{\overline{D\times[0,T)}}u$,
then we are ensured that $u\equiv  u(x_0,y_0)$ in $D\times[0,T)$.
Bearing in mind that the value of the field on the boundary of the
invisibility region $D$ in Figures \ref{fig:mimetism7b}-\ref{fig:mimetism7a} is the solution
of the transformed equation (\ref{transfpressure1}), we note that it
is an appromation of the field
solution of the diffusion equation  (\ref{govpressure}) at the
point $r=0$ in (\ref{magictransfo}) mapped onto $r'\in\partial D$.
We therefore deduce that the field
in the circular invisibility region in Figures \ref{fig:mimetism7b}-\ref{fig:mimetism7a}
is a constant field equal
to the value of the field at the center point of the
disc $D$ before geometric transform. In the present
case, the value of $u$ solution of (\ref{govpressure})
at the center point is $T/2$, hence after the
geometric transform the value of $u$
solution of (\ref{transfpressure1})
on $\partial D$ equals $T/2$, and so is the
value of $u_0$ solution of its
approximate
homogenized equation (\ref{acoustic-hom2}).
Applying the above
corollary of the mean value
theorem, we deduce that
$u\equiv T/2$ in the disc $D$.
Importantly, when we consider a
different point as the origin of the
geometric transform, the
maximum value inside $D$
can be smaller or larger than
$T/2$, but it cannot exceed
$T$. Indeed, if we assume that
the value of the field
at the point taken as the
origin $r=0$ of the geometric
transform is $T$, maximum of the
field
solution of (\ref{govpressure}), then
the above discussion leads
to $u\equiv T$ inside the
invisibility region.


\section{Conclusion and perspectives}
In this review article, we have proposed some models of generalized cloaks that create
some optical illusion for light in diffusive media: infinite conducting obstacles dressed with these cloaks
scatter light like other infinite conducting obstacles. In these cloaks,
an electric wire could in fact be hiding a larger object near it. In fact, any object
could mimic the signature of another.

\begin{figure}[h]
\hspace{3cm}\mbox{}\scalebox{0.5}{
\includegraphics[width=20cm,angle=0]{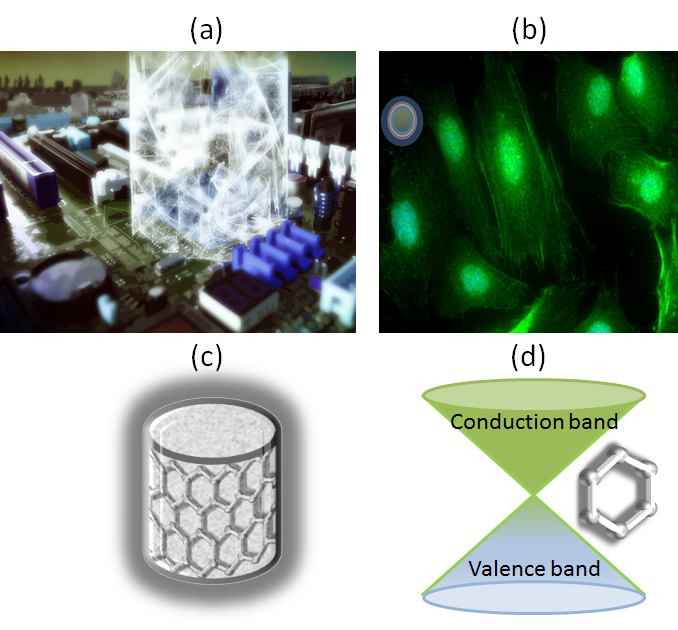}}
\caption{Artistic view of a cylindrical diffusion cloak enveloping components of a computer motherboard (a), a biocloak in a living organism (b), a layer of graphene conformally mapped on a cylinder (c) and graphene's electronic band structure with a Dirac cone (d).
Note that the high diffusivity of graphene (induced by its unique electronic band structure without gap between conduction and valence band) makes a good candidate for the inner most layer of the layered diffusion cloak, that would also consist of an alternation of peptides of markedly different diffusivities as proposed in \cite{guenneau13}.
}
\label{fig:intro}
\end{figure}

For instance, we design a cylindrical cloak so that a circular obstacle behaves like a square obstacle, thereby bringing about one of the oldest
enigma of ancient time : squaring the circle! The ordinary cloaks then come as a particular
case, whereby objects appear as an infinitely small infinite conducting object
(of vanishing scattering cross-section) and hence
become invisible. Such generalized cloaks are described by nonsingular permittivity and
permeability, even at the cloak's inner surface. We
have proposed and discussed some interesting applications. The results within this paper
can be extended to cloaking and anamorphism for electromagnetic, acoustic, hydrodynamic and elastodynamic waves
propagating in anisotropic heterogeneous media.

Regarding the fabrication of the light diffusion cloak, there is of course the route
work \cite{wegener14} but we note the possibility to use instead
a self-assembly technique for colloidal nanoparticles like in \cite{mulhig13}. For
the mass diffusion cloak, it has been shown in \cite{puvirajesinghe16} that
certain membranes of graphene oxyde allow for control of drug diffusion,
we thus believe GO could be used as a building block of a biocloak
(see Fig. \ref{fig:intro} for an artistic view) that would
also consist of an alternation of peptides as proposed in \cite{guenneau13}.
Indeed, a cloak whose innermost layer is made of graphene would meet the
requirement of high anisotropy of transformed diffusivity in (\ref{rhort1a}).

Let us conclude this review by some perspectives on diffusion media mimicking
celestial mechanics \cite{black0,black1,black2,black3,black4,black5}, conformal
mapping for diffusion processes \cite{ulf} and some governing
equations in other fields of physics.

\subsection{Black hole analogues for heat, light and other diffusion phenomena}
Concept of black hole was first discussed by Laplace in the framework of the Newton mechanics, whose event
horizon radius is the same as the Schwarzschild's solution of the Einstein's vacuum field equations. If all
those objects having such an event horizon radius but different gravitational fields are called as black holes,
then one can simulate certain properties of the black holes using electromagnetic fields and metamaterials
due to the similar propagation behaviours of electromagnetic waves
in curved space and in inhomogeneous
metamaterials. In a recent theoretical work by Narimanov and Kildishev \cite{black0}, with an independent
proposal by Cheng and Cui \cite{black1}, an optical black hole has been
proposed based on metamaterials, in which the theoretical analysis
and numerical simulations showed that
all electromagnetic waves hitting it are trapped and absorbed. 
The  Schwarzschild metric of a black hole in cylindrical coordinates
$(r,\theta)$ is
\begin{equation}
ds^2=\left( 1-L/r \right) c^2 dt^2 - \frac{1}{1-L/r} dr^2
-r^2 d\theta^2 - dz^2 \; ,
\end{equation}
and in spherical coordinates
$(r,\theta,\Theta)$ it is
\begin{equation}
ds^2=\left( 1-L/r \right) c^2 dt^2 - \frac{1}{1-L/r} dr^2
-r^2 \left( d\theta^2+\sin^2\theta d\Theta^2 \right) \; ,
\end{equation}
where $L$ is the event horizon of the black hole.

If we are interested in control of heat, transformation
thermodynamics \cite{guenneau12},
we get that a
medium with anisotropic
heterogeneous conductivity
and heterogeneous
denisty and heat capacity
\begin{equation}
\underline{\underline{\kappa}}
=
\frac{1}{1-L/r}
\left(
\begin{array}{cc}
1-\frac{L}{r}\frac{x^2}{r^2} & -\frac{L}{r}\frac{xy}{r^2} \nonumber \\
-\frac{L}{r}\frac{xy}{r^2} & 1-\frac{L}{r}\frac{y^2}{r^2}  
\end{array}
\right)
\; , \;
\hbox{ and }
\rho c = \frac{1}{1-L/r} \; ,
\end{equation}
where $r=\sqrt{x^2+y^2}$,
should be the analogue of
a black hole for heat.

We obtain the following transformed conductivity and
product of density and heat capacity in cylindrical
coordinates
\begin{equation}
\underline{\underline{\kappa}}
=\rm{\bf{Diag}}\left(1,\frac{1}{1-L/r} \right)
\; , \; \hbox{ and }
\rho c
=\frac{1}{1-L/r}
\; ,
\end{equation}
and in spherical coordinates
\begin{equation}
\underline{\underline{\kappa}}
=\rm{\bf{Diag}}\left(1,\frac{1}{1-L/r},\frac{1}{1-L/r} \right)
\; , \; \hbox{ and }
\rho c
=\frac{1}{1-L/r}
\; ,
\end{equation}
where $L$ is the event horizon.

It is of course possible to translate this in the language of transformation
light diffusion, or mass diffusion, and this is left as an exercise. We
further note that other designs of optical black holes \cite{black3,black4,black5}
and wormholes \cite{black6,black7} can be applied to diffusion process.


\subsection{Conformal mappings for light, heat, mass and other diffusion physics}
Inspired by the seminal paper of Leonhardt on conformal optics \cite{ulf}, we consider a homogeneous isotropic medium within which the following governing equation for a diffusion process holds: 
\begin{equation}
\left(\frac{\partial^2}{\partial x^2}+\frac{\partial^2}{\partial y^2}+\Omega\right)
u=0,
\label{eq:helmholtz}
\end{equation}
for $u(x,y)$ in an infinite transverse plane $-\infty<x,y<\infty$. Note that for instance (\ref{govpressure}) can be
encapsulated in (\ref{eq:helmholtz}), simply by taking $\Omega=D^{-1}(i\omega\nu^{-1}-\mu_a)$. In practice, the metamaterial will have a
finite extension and the diffusion direction of pseudo waves may be slightly tilted without causing an
appreciable difference to the ideal case (note that unlike for \cite{ulf}, $\Omega$ lies in the complex plane, not on the positive real line).

To describe the spatial coordinates in the
propagation plane, complex numbers $z=x+iy$
are used with the partial derivatives $\partial_x=\partial_z+\partial_z^\star$
and $\partial_y=i\partial_z-i\partial_z^\star$,
where the asterisk symbolizes
complex conjugation. In this
way, the Laplace operator takes the form $\partial^2_x+\partial^2_y=4\partial^\star_z\partial_z$
so that (\ref{eq:helmholtz}) is expressed in complex coordinates as follows
\begin{equation}
\left(4{\left(\frac{\partial}{\partial z}\right)}^\star\frac{\partial}{\partial z}+\Omega\right)
u=0 \; .
\label{eq:helmholtz1}
\end{equation}
In the case of gradually varying diffusivity,
that is $\Omega:=\Omega(z)$, one needs to assume that $\Omega$
must not vary by much over the scale of a diffusion length.

Suppose we introduce new coordinates $w$ described by an
analytic function $w(z)$ that does not depend on
$z^\star$. Such functions define conformal maps \cite{nehari}
that preserve the angles between the coordinate
lines. Because
\begin{equation}
{\left(\frac{\partial}{\partial z}\right)}^\star\frac{\partial}{\partial z}
={\mid \frac{dw}{dz}\mid}^2
{\left(\frac{\partial}{\partial w}\right)}^\star\frac{\partial}{\partial w} \; ,
\label{newomega}
\end{equation}
we obtain in $w$ space
a diffusion equation with the
transformed index profile $\Omega'(w)$ that is
related to the original one as
\begin{equation}
\left(4{\left(\frac{\partial}{\partial w}\right)}^\star\frac{\partial}{\partial w}+\Omega'\right)
u'=0,
\label{eq:helmholtz2}
\end{equation}

Suppose that the diffusive medium is designed such that
$\Omega(z)$ is the modulus of an analytic function $g(z)$.
The integral of $g(z)$ defines a map $w(z)$ to new
coordinates where, according to (\ref{newomega}), the
transformed index $\Omega'$ is unity. Consequently, in
$w$ coordinates, the diffusion is indistinguishable
from empty space where diffusion flux follows straight lines. The medium
performs a diffusion conformal mapping to
homogeneous space. If $w(z)$ approaches $z$ for $w\to\infty$,
all incident pseudo-waves appear at infinity as if they
have diffused through empty space, regardless
of what has happened in the medium. However,
as a consequence of the Riemann Mapping
Theorem \cite{nehari}, nontrivial $w$ coordinates occupy
Riemann sheets with several $\infty$, one on each
sheet.

Following \cite{ulf}, we consider the simple map
\begin{equation}
w=z+\frac{a^2}{z} \; , \; z=\frac{1}{2}+\left( w\pm\sqrt{w^2-4a^2} \right) \; ,
\label{necloak}
\end{equation}
that is realized by the pseudo refractive index profile $\Omega=\mid 1-a^2/z^2\mid$. The
constant $a$ characterizes the spatial extension of
the medium. The function (\ref{necloak}) maps the exterior
of a circle of radius $a$ on the $z$ plane onto a given
Riemann sheet and the interior onto another.
Pseudo-waves traveling on the exterior $w$ sheet may
well be passing through the branch cut
between the two branch points $\pm 2a$. In
continuing their diffusion, the pseudo-waves approach
$\infty$ on the interior $w$ sheet. From the point of view of
an observer on the physical $z$
plane, they cross the circle of radius $a$ and
approach the singularity of the pseudo refractive index
at the origin. For general $w(z)$, only one $\infty$ on
the Riemann structure in $w$ space corresponds
to the true $\infty$ of physical $z$ space and the
others to singularities of $w(z)$. Instead of
traversing space, pseudo-waves may cross the
branch cut to another Riemann sheet where
they approach $\infty$. For an observer in physical space, the
rays seem to be attracted towards some
singularities of the pseudo refractive index. Instead
of becoming invisible, the medium casts a
shadow that is as wide as the apparent size of
the branch cut. Nevertheless, the diffusion on
Riemann sheets could serve as a
mathematical tool for developing the
design of metamaterials that shape the flow of diffusion processes.

\begin{figure}[h]
\hspace{1cm}\mbox{}\scalebox{0.7}{
\includegraphics[width=20cm,angle=0]{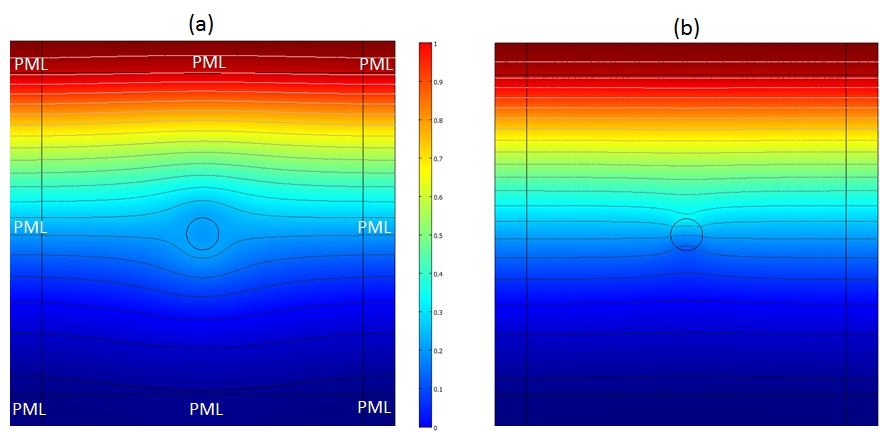}}
\caption{Diffusive light scattering at angular frequency $\omega=\pi/10$ rad.s$^{-1}$ within a conformal cloak given by (\ref{necloak}) of infinite extent with an inner boundary of radius $a=0.2m$ hosting an object with diffusivity ten times larger than that of cloak at $z=0.25$m (a) and scattering same object in a homogeneous diffusive medium with homogeneous diffusivity like in (a) at $z=0.25$m. One notes the streamlines are detoured around the object in (a), whereas they are attracted by the same object in (b). The cloak acts as a repeller for objects.

}
\label{conformal}
\end{figure}

We illustrate numerically this conformal cloak for diffusion processes in Fig. \ref{conformal}. Results of finite element computations are shown in Fig. \ref{conformal} for a conformal cloak given by (\ref{necloak}) for $a=0.25$m and a finite angular frequency of $\omega=\pi/10$ rad.s$^{-1}$ in panel (a) where we note that we used our specially designed PMLs as detailed in section \ref{pmlheat}, but here the background medium is spatially varying, and so are PMLs. In panel (b), the object illuminated by the same pseudo wave in a homogeneous diffusive medium acts as a concentrator, the streamlines are pinched inside it (so the flux therein is large).

\subsection{Non-Eudlidean cloaking for diffusion processes}
In the sequel, we make use of the correspondence between the anisotropic diffusion equation
for heat (\ref{heat1}) and the anisotropic equations (\ref{eq:El1}) and (\ref{eq:El1}) for transverse
electromagnetic waves.
In this paper, we focus on a 2D version of the non-Euclidean
invisibility cloak that has been described in the previous sections in flat space.
We perform the earlier described geometric transformation from
2D virtual space to the plane $xy$ of physical space, and then,
we add an extra dimension $z$, assuming that the material parameters
only depend upon $xy$ coordinates, thereby generating
a cylindrical 3D physical space. We consider transverse electromagnetic waves,
which split in $s$ and $p$ polarizations.
For calculating the line elements in this 3D physical space, we then
have to add the term $dz^2$ to the line elements corresponding to
the 2D situation.
To describe the Euclidean part of the cloak, we use bipolar
cylindrical coordinates $(\tau,\sigma,z)$ in physical space that are related
to the Cartesian $(x, y, z)$ as
\begin{equation}
x=\frac{a\sinh\tau}{\cosh\tau-\cos\sigma} \; , \; y=\frac{a\sin\sigma}{\cosh\tau-\cos\sigma} \; , \; z=z \; .
\label{tyc1}
\end{equation}
Here, the parameter $a$ characterizes the size of the cloak. Analogous
relations hold in virtual space where $x$, $y$, and $\sigma$ are
replaced by $x'$, $y'$, and $\sigma'$, respectively. The mapping between
the two spaces is given in the bipolar cylindrical coordinates by
the following relation between $\sigma$ and $\sigma'$:
\begin{equation}
\sigma'=\sigma \; , \; \rm{for} \mid\sigma\mid\leq\pi/2 \; ,
\label{tyc2}
\end{equation}

and

\begin{equation}
\sigma'=\left(\frac{4\sigma^2}{\pi} - 3\mid\sigma\mid + \pi \right) \; , \; \rm{for} \frac{\pi}{2}<\mid\sigma\mid<\frac{3\pi}{4} \; ,
\label{tyc3}
\end{equation}
and the coordinates $\tau$, $z$ coincide in the two spaces (the mapping only involves inplane coordinates). The line
element $ds^2 = d{x'}^2 +d{y'}^2 +d{z'}^2$ in virtual space is equal to
\begin{equation}
ds^2=\frac{a^2}{{(\cosh\tau-\cos\sigma')}^2} (d\tau^2+d{\sigma'}^2)+dz^2 \; .
\label{tyc4}
\end{equation}

The tensors of permittivity and permeability for this
region have been calculated by Tyc et al. in \cite{tyc10} using the same general method
as in \cite{tyc09a}, see also the supplemental material \cite{tyc09b}. A short, but
insightful, review of \cite{tyc09a} has been written by experts in generalized
cloaking \cite{nicolet09a} and can serve as an entry door in this field. In
\cite{tyc10}, the following expressions can be found:
\begin{equation}
\underline{\underline{\varepsilon}}=\underline{\underline{\mu}}= \rm{Diag} \left( \frac{d\sigma}{d\sigma'},\frac{d\sigma'}{d\sigma},\frac{(\cosh\tau-\cos\sigma)}{(\cosh\tau-\cos\sigma')}\frac{d\sigma'}{d\sigma}\right) \; ,
\label{tyc5}
\end{equation}
which can be translated in terms of anisotropic diffusivity and product of density and heat capacity in the anisotropic governing equation for heat, see Eq. (\ref{heat1}):
\begin{equation}
\underline{\underline{\kappa}}=\rm{Diag} \left( \frac{d\sigma}{d\sigma'},\frac{d\sigma'}{d\sigma} \right)
\rm{ and } \; \rho c=\frac{(\cosh\tau-\cos\sigma)}{(\cosh\tau-\cos\sigma')}\frac{d\sigma'}{d\sigma} \; .
\label{tyc6}
\end{equation}

The mapping between the non-Euclidean part of virtual space
and the corresponding part of physical space requires more work.
Here, we will not reproduce all the corresponding formulas of
\cite{tyc09b}, but give some important steps, as was done in \cite{tyc10}. One
starts with a sphere of radius $r=(4/\pi) a$
(see Eq. (S43) in \cite{tyc09b}) parameterized by spherical coordinates—
the latitudinal angle $\theta$ and the longitudinal angle that is denoted
by $\sigma'$. This is not yet the sphere of virtual space, but is related
to it by a suitable M\" obius transformation (see Eq. (S40) in \cite{tyc09b}) represented
on the sphere via stereographic projection (see Eq. (S37)-(S39) in \cite{tyc09b}).
Then, the coordinates $(\sigma',\theta)$ are mapped to the
bipolar coordinates $(\sigma,\tau)$ of physical space as follows. The
mapping between $\sigma'$ and $\sigma$ is given by (\ref{tyc3}),
where $\sigma'$ is set to
one of the intervals $[-2\pi,-\pi]$ or $[\pi, 2\pi]$ by adding or subtracting
$2\pi$. The corresponding intervals of $\sigma$ are then $[-\pi,-3\pi/4]$
and $[3\pi/4, \pi]$. The mapping between $\theta$ and $\tau$ can be derived by
combining Eqs. (S41) and (S46) in \cite{tyc09b}. After some calculations,
this yields
\begin{equation}
\theta=2\arctan(t-1+\sqrt{t^2+1} \; , \; \rm{ with } \; t=\tan\left( \frac{\pi}{4} \left( \frac{\sinh\tau}{\cosh\tau+1}\right) + \frac{\pi}{4} \right) \; .
\label{tyc7}
\end{equation}

The line element of the non-Euclidean part of virtual space is

\begin{equation}
ds^2= \frac{16a^2}{\pi^2} {\left( \frac{1+\cot^2(\theta/2)}{1-2\cos\sigma'\cot(\theta/2)+2\cot^2(\theta/2)}\right)}^2 (d\theta^2+\sin^2\theta d{\sigma'}^2)+dz^2 \; .
\label{tyc8}
\end{equation}

This line element is slightly different from the line element in Eq. (S47) of \cite{tyc09b}, as noted in \cite{tyc10}, where details can be found.

Tyc et al. simplify Eq. (\ref{tyc8}) in a more amenable form in \cite{tyc10} that leads to the following entries for the tensors of permittivity and permeability
\begin{equation}
\begin{array}{ll}
\underline{\underline{\varepsilon}} & =\underline{\underline{\mu}}= \rm{Diag} \left( \varepsilon_\sigma, \varepsilon_\tau, \varepsilon_z \right) \; ,
\hbox{ with } \\
%
\varepsilon_\sigma & = \displaystyle{\frac{\pi}{4}\frac{(t^2+1+t\sqrt{t^2+1})}{{({t-1+\sqrt{t^2+1}})}} \frac{1}{(\cosh\tau+1)}\frac{d\sigma}{d\sigma'}} \\
\varepsilon_\tau &= \displaystyle{\frac{1}{\varepsilon_\sigma}}\\
\varepsilon_z &= \displaystyle{\frac{16}{\pi}\frac{(t-1+\sqrt{t^2+1}) (t^2+1+t\sqrt{t^2+1})}{{(1+{(t-1+\sqrt{t^2+1})}^2)}^2}
\frac{u^2 {(\cosh\tau-\cos\sigma)}^2}{(\cosh\tau+1)}\frac{d\sigma'}{d\sigma}} \; .
\end{array}
\label{tyc10}
\end{equation}

Regarding the heat equation, this translates as follows in terms of anisotropic diffusivity and product of density by heat capacity
\begin{equation}
\underline{\underline{\kappa}}=\rm{Diag} \left( \varepsilon_\sigma, \varepsilon_\tau \right)
\rm{ and } \; \rho c= \varepsilon_z \; .
\label{tyc6}
\end{equation}
One can easily adapt these parameters to the DPDW and mass diffusion equations and this is left as a simple exercise.

Interestingly, Smerlak derived the Fokker-Plank equation in the general curved space-time framework back in 2012 \cite{smerlak}.
Bearing this in mind, we would like to end this review article by some perspectives on non-Eudlidean cloaking for diffusion processes
in curved spaces.

\subsection{Other transformation based governing equations}
There is a vast literature on diffusion equations, that span chemical, medical, physical and engineering sciences, and that also appear in economics and image processing. There is therefore room for further investigations of transformation based diffusion equations applied to societal problems where anisotropy might play a more prominent role.
Since we have a particular interest in mass diffusion applied to biocloaks, we first consider a diffusion model with a mathematical setup by Haidar \cite{haidar} for a neutron wave propagation akin to heat pseudo-waves, motivated by modulated neutron source experiments with applications in Boron and Gadolinium capture for cancer treatment \cite{hosmane}. Let $\phi$ denote the neutron flux distribution in a Boron and/or Gadolinium loaded cancerous tissue of thin slab domain $\Omega$, the diffusion equation of a flux of speed $\nu$ thermal neutrons in $\Omega$ is
similar to that of diffusive light (\ref{govpressuret}), more precisely 
\begin{equation}
-\nabla\cdot \left(D({\bf x}) \nabla \Phi({\bf x},t)\right) + \frac{1}{\nu({\bf x})}\frac{\partial\Phi({\bf x},t)}{\partial t} + \mu_{a}({\bf x}) \Phi({\bf x},t) = S_0({\bf x},t) \; . \label{govpressuretneutron}
\end{equation}
where $\phi({\bf x},t)=\nu N({\bf x},t)$, with $N$ the neutron density. Moreover, $D$ is the diffusion length of these neutrons in $\Omega$, $\mu_a$ is the macroscopic absorption cross section of these neutrons and $D/\mu_a=L^2$, where $L$ is
the diffusion length of these neutrons. In light water, $D=0.164$ cm, $\mu_a=0.0198$ cm$^{-1}$, and $L=2.88$ cm. The life time of thermal neutrons is $T={(\nu\mu_a)}^{-1}=213$ $\mu$s in $H_2O$. The source $S_0$ is periodic, and boundary
conditions on $\partial\Omega$ can be zero flux $\phi({\bf x},t)\mid\partial\Omega=0$ (Dirichlet), modulated neutron flux intensity $[D({\bf x}) \nabla \Phi({\bf x},t)] \cdot {\bf n}=\xi S_0$ (Neumann) where $\xi$ is a coupling parameter between $\Omega$ and an adjacent domain, which is filled for instance with non cancerous cell.

Another equation of interest is the Fokker-Planck equation, that describes the time evolution of the probability density function of the velocity of a particle under the influence of drag forces and random forces, as in Brownian motion.
For an Ito process driven by the standard Wiener process ${\bf W}_t$ and described by the stochastic differential equation $d{\bf X}_t=\boldsymbol{\mu}({\bf X}_t,t)dt+\boldsymbol{\sigma}({\bf X}_t,t)d{\bf W}_t$ where ${\bf X}_t$ and ${\displaystyle {\boldsymbol {\mu }}(\mathbf {X} _{t},t)}$ are N-dimensional random vectors, ${\displaystyle {\boldsymbol {\sigma }}(\mathbf {X} _{t},t)} {\boldsymbol {\sigma }}(\mathbf {X} _{t},t)$ is an $N {\displaystyle \times } M$ matrix and ${\displaystyle \mathbf {W} _{t}}$ is a $M$-dimensional standard Wiener process, the probability density ${\displaystyle p(\mathbf {x} ,t)}$ of the random variable ${\displaystyle \mathbf {X} _{t}}$ satisfies the Fokker-Planck equation \cite{kadanoff}
\begin{equation}
\frac{1}{2}\sum_{i=1}^N \sum_{j=1}^N  \frac{\partial^2}{\partial x_i\partial x_j}\left(D_{ij}({\bf x},t) p({\bf x},t)\right) - \sum_{i=1}^N  \frac{\partial}{\partial x_i}\left(\mu_{i}({\bf x},t) p({\bf x},t)\right) =\frac{\partial p({\bf x},t)}{\partial t} \; , \label{govpressuretfk}
\end{equation}
with drift vector ${\displaystyle {\boldsymbol {\mu }}=(\mu _{1},\ldots ,\mu _{N})}$ and diffusion tensor $D_{ij}({\bf x},t)=\sum_{k=1}^M\sigma_{ik}({\bf x},t)\sigma_{jk}({\bf x},t)$.

Under coordinate change ${\bf x}\longmapsto {\bf x}'$, this equation takes the form  
\begin{equation}
\begin{array}{cc}
&\displaystyle{\frac{1}{2}\sum_{i=1}^N \sum_{j=1}^N\sum_{k=1}^N \sum_{l=1}^N  \frac{\partial}{\partial x'_i}\rm{det}J_{ij}\left(J_{ik}^{-1}D_{kl}({\bf x}',t) J_{jl}^{-T}\frac{\partial}{\partial x'_j} p({\bf x}',t)\right)} \nonumber \\
&- \displaystyle{\sum_{i=1}^N\sum_{j=1}^N  (\rm{det}J_{ij}) J_{ji}^{-T}\frac{\partial}{\partial x'_i} \left(\mu_{i}({\bf x}',t) p({\bf x}',t)\right) = \rm{det}J_{ij}\frac{\partial p({\bf x}',t)}{\partial t}} \; ,
\end{array}
\label{govpressuretfkbis}
\end{equation}
with ${\bf J}=\partial (x_1,...,x_N)/\partial(x'_1,...,x'_N)$ the Jacobian, ${\bf J}^{-1}$ its inverse and ${\bf J}^T$ its transpose.

Other PDEs of interest for coordinate changes include the Boltzmann transport equation \cite{boltzmann,harris71,villani98}, which thanks to Mouhot and Villani \cite{villani11} is know to have well-behaved solutions (if a system obeying the Boltzmann equation is perturbed, then it will return to equilibrium, rather than diverging to infinity).
In a simplified form, it can be written similarly to the Vlasov-Poisson nonlinear equation (also known as Landau equation)
\begin{equation}
{\bf v} \cdot \nabla_{\bf x} \Phi({\bf x},t,{\bf v}) + \frac{\bf F}{m} \cdot\nabla_{\bf v} \Phi({\bf x},t,{\bf v}) + \frac{\partial\Phi({\bf x},t,{\bf v})}{\partial t} =0 \; .
\label{villani1}
\end{equation}
where the unknown function, $\Phi$ is the density of electrons in a gas in the phase space of all positions and velocities of particles (${\bf x},{\bf v}$), $m$ is the mass of an electron and the mean field electrostatic force ${\bf F}=-e{\bf E}=-e\nabla\Delta^{-1}(4\pi\rho)$, with $e=1.6021766208 \times 10^{-19}$ Coulomb and $\rho({\bf x},t)$ the density of charges.
All quantities have $N\geq 2$ indices.

Under coordinate change ${\bf x}\longmapsto {\bf x}'$, assuming that ${\bf v'}({\bf x}')={\bf v}({\bf x}')$, the Vlasov-Poisson nonlinear equation (\ref{villani1}) takes the form 
\begin{equation}
\rm{det}{\bf J} \left( {\bf v} \cdot {\bf J}^{T}\nabla_{\bf x'} \Phi({\bf x}',t,{\bf v}) + \frac{\bf F}{m} \cdot\nabla_{\bf v} \Phi({\bf x}',t,{\bf v}) + \frac{\partial\Phi({\bf x},t,{\bf v}')}{\partial t} \right) =0 \; ,
\label{villani1bis}
\end{equation} 
with ${\bf J}=\partial (x_1,...,x_N)/\partial(x'_1,...,x'_N)$.

Finally, we would like to mention that geometric transforms might find also interesting applications in mathematical models for market. For instance, the Black-Scholes equation \cite{blackscholes1,blackscholes2}.
Let $s_i(t)\in [0,+\infty)$, $i = 1, 2, . . . , N$ denote the value of the $i^{th}$ underlying asset at time $t>0$ and $u({\bf s},t)$ denote the price of an option, where ${\bf s} = (s_1, s_2, . . . , s_N)$.
Then the option price follows the following generalized Black–Scholes partial differential equation
\begin{equation}
-\frac{1}{2}\sum_{i=1}^N \sum_{j=1}^N  \rho_{ij}\sigma_i\sigma_j s_i s_j \frac{\partial^2}{\partial s_i\partial s_j} u({\bf s},t) - \sum_{i=1}^N  r s_i\frac{\partial}{\partial s_i} u({\bf s},t) + r u({\bf s},t) =\frac{\partial u({\bf s},t)}{\partial t} \; ,
\label{blackscholes1}
\end{equation}
with final condition $u({\bf s},T) = \Lambda({\bf s})$ (payoff function at maturity $T$), where $r$ is the constant riskless interest rate, $\sigma_i$ are volatilities of $s_i$, and $\rho_{ij}$ are the asset correlations between $s_i$ and $s_j$.

Under coordinate change ${\bf s}\longmapsto {\bf s}'$, this equation takes the form  
\begin{equation}
\begin{array}{cc}
&\displaystyle{-\frac{1}{2}\sum_{i=1}^N \sum_{j=1}^N\sum_{k=1}^N \sum_{l=1}^N  \frac{\partial}{\partial s'_i}\rm{det}J_{ij}\left(J_{ik}^{-1}\rho_{kl}\sigma_k\sigma_l s'_k s'_l J_{jl}^{-T}\frac{\partial}{\partial s'_j} u({\bf s}',t)\right)} \nonumber \\
&- \displaystyle{\sum_{i=1}^N\sum_{j=1}^N  (\rm{det}J_{ij}) J_{ji}^{-T} r s'_i\frac{\partial}{\partial s'_i} u({\bf s}',t) + \rm{det}J_{ij} r u({\bf s}',t) = \rm{det}J_{ij}\frac{\partial u({\bf s}',t)}{\partial t}} \; ,
\end{array}
\label{govpressuretfkbis}
\end{equation}
with ${\bf J}=\partial (s_1,...,s_N)/\partial(s'_1,...,s'_N)$ the Jacobian, ${\bf J}^{-1}$ its inverse and ${\bf J}^T$ its transpose. This transformed Black-Scholes equation might find some applications in control of options trading.
The transformed Black-Scholes model could help hedge the option by buying and selling the underlying asset in just the right way and, as a consequence, to eliminate risk. This could serve as the basis of hedging strategies
such as those engaged in by investment banks and hedge funds, by adding anisotropy in the existing models.

We would like to finalise this review article by some perspectives on image processing. Back in 2009, some of us have applied a simple algorithm for comparative analysis of different datasets related to heparan sulphate oligosaccharides \cite{image1}. In this work, it was observed that when a cloud of points displays some preferential direction (anisotropy) for repeated experiments, the quantification of their reproducibility can be made through evaluation of an average Lyapunov exponent characterizing the area-preserving nature of a sequence of effective ellipses \cite{image1}. We believe that the evolution of the data sets with time could be encapsulated in a diffusion equation with anisotropic coefficients. It is actually well-known
that diffusion of heat smoothes the temperature function $u$, and the noting that $u(x,t)$ is equivalently found by minimizing problems of $L^2$ norm of gradients using the identity $\partial u/\partial t=\nabla\cdot(\nabla u/{\Vert u \Vert})$, some denoising algorithm can be applied to images thanks to anisotropic diffusion equation \cite{image2,image3,image4}. We believe that image processing could benefit from geometric transforms so as to improve smoothing algorithms.

\section*{Acknowledgements}
A.D. and S.G. acknowledge funding from ERC grant ANAMORPHISM. S.G. T.P. acknowledge funding
from AMIDEX foundation of Aix-Marseille University through project BIOCLOAK. M.F. acknowledges funding from the Qatar National Research Fund (QNRF) through a National Priorities Research Program (NPRP) Exceptional grant, NPRP X-107-1- 027.

\end{document}